\documentclass[a4paper,11pt]{article}
\pdfoutput=1 

\usepackage{jheppub}

\usepackage[T1]{fontenc}

\usepackage{amsmath,amssymb,amsfonts, physics}
\usepackage{mathrsfs}
\usepackage{bbm}
\usepackage{slashed}
\usepackage{graphicx}
\usepackage{verbatim}
\usepackage[T1]{fontenc}
\usepackage[utf8]{inputenc}
\usepackage[dvipsnames]{xcolor}
\usepackage[normalem]{ulem}
\usepackage{svg}

\usepackage[backend=bibtex,style=numeric-comp, sorting=none]{biblatex}
\bibliography{references}

\definecolor{pink}{RGB}{255,192,203}

\title{\boldmath Formation and evaporation of quantum black holes from the decoupling mechanism in quantum gravity}

\author[a,b]{Johanna N.~Borissova}
\author[a,c]{Alessia Platania}
\affiliation[a]{Perimeter Institute for Theoretical Physics, 31 Caroline Street North, Waterloo, ON N2L 2Y5, Canada}
\affiliation[b]{Department of Physics and Astronomy,  University of Waterloo, 200 University Avenue West, Waterloo, ON N2L 3G1, Canada}
\affiliation[c]{Nordita, KTH Royal Institute of Technology and Stockholm University, Hannes Alfv\'ens v\"ag 12, SE-106 91 Stockholm, Sweden}

\emailAdd{jborissova@perimeterinstitute.ca}
\emailAdd{aplatania@perimeterinstitute.ca}

\abstract{We propose a new method to account for quantum-gravitational effects in cosmological and black hole spacetimes. At the core of our construction is the ``decoupling mechanism'': when a physical infrared scale overcomes the effect of the regulator implementing the Wilsonian integration of fluctuating modes, the renormalization group flow of the scale-dependent effective action freezes out, so that at the decoupling scale the latter approximates the standard quantum effective action. Identifying the decoupling scale allows to access terms in the effective action that were not part of the original truncation and thus to study leading-order quantum corrections to field equations and their solutions. Starting from the Einstein-Hilbert truncation, we exploit for the first time the decoupling mechanism in quantum gravity to investigate the dynamics of quantum-corrected black holes from formation to evaporation. Our findings are in qualitative agreement with previous results in the context of renormalization group improved black holes, but additionally feature novel properties reminiscent of higher-derivative operators with specific non-local form factors.}

\begin{document} 
\maketitle
\flushbottom

\section{Introduction}\label{sect:introduction}

The observation of gravitational waves~\cite{LIGOScientific:2016aoc} and the reconstruction of the image of a black hole shadow~\cite{EventHorizonTelescope:2019dse} have provided impressive support to Einstein's General Relativity (GR), and to the existence of astrophysical objects whose properties reflect very closely those of GR's black holes.

Yet, it is expected that GR eventually breaks down at Planckian scales, leaving the stage to a more fundamental theory accounting for the quantum nature of gravity. Among the different theories of quantum gravity, string theory~\cite{Berkovits:2022ivl}, loop quantum gravity~\cite{Ashtekar:2021kfp} and spin foams~\cite{Perez:2012wv}, asymptotically safe gravity~\cite{Percacci:2017fkn,Reuter:2019byg}, and non-local gravity~\cite{Buoninfante:2016iuf,Modesto:2017sdr} have gained considerable attention. While seemingly diverse, a common feature is that the effective action and field equations stemming from their ultraviolet (UV) completions display additional higher-derivative terms~\cite{Gross:1986mw,Grimm:2004uq,Modesto:2013ioa,Hohm:2014sxa,Hohm:2016lim,Donoghue:2021cza,Buoninfante:2018mre,Knorr:2019atm,Mayer:2020lpa,Borissova:2022clg} which complement the Einstein-Hilbert dynamics. These corrections are expected to play an important role in determining the quantum spacetimes allowed by a principle of least action~\cite{Knorr:2022kqp} and their dynamics. In this respect, black holes and their alternatives are particularly important avenues: on the one hand, quantum gravity is expected to yield non-singular solutions~\cite{Ayon-Beato:1999qin,Bonanno:2000ep,Bronnikov:2000vy,Dymnikova:2004zc,Modesto:2004xx,Hayward:2005gi,Ansoldi:2006vg,Modesto:2008im,Ansoldi:2008jw,Nicolini:2008aj,Hossenfelder:2009fc,Modesto:2010rv,Spallucci:2011rn,Sprenger:2012uc,Bambi:2013caa,Culetu:2014lca,Frolov:2014jva,Casadio:2014pia,Carr:2015nqa,Frolov:2016pav,Bonanno:2016dyv,Bonanno:2017kta,Bonanno:2017zen,Adeifeoba:2018ydh,Buoninfante:2018stt,Carballo-Rubio:2018jzw,Carballo-Rubio:2019fnb,Simpson:2019mud,Platania:2019kyx,Bonanno:2019ilz,Bosma:2019aiu,Carballo-Rubio:2019nel,Lan:2020fmn,Lobo:2020ffi,Franzin:2021vnj,Maeda:2021jdc,Bokulic:2022cyk,Cadoni:2022chn,Casadio:2022ndh} or spacetimes with integrable singularities~\cite{Lukash:2011hd,Lukash:2013ts}; on the other hand, the quantum dynamics could shed light on how these objects are formed in a gravitational collapse, what the final stages of the evaporation process could be, and how singularity resolution can be achieved in the gravitational path integral~\cite{Borissova:2020knn}. Moreover, accounting for the  quantum dynamics is key to establish whether the linear instabilities that potentially affect the inner horizon of regular or rotating black holes~\cite{DiFilippo:2022qkl,Bonanno:2020fgp} are damped or enhanced by quantum effects. Finally, the number of derivatives in the effective action is crucially related to the type of allowed solutions: truncating the full effective action to quartic order in a derivative expansion, the phase space of all possible solutions is dominated by wormholes and singular black holes~\cite{Lu:2015cqa,Lu:2015tle,Lu:2015psa,Bonanno:2019rsq,Bonanno:2022ibv}. Adding terms with six or more derivatives, the field equations instead admit spherically symmetric regular solutions~\cite{Holdom:2002xy,Knorr:2022kqp}. 

Determining the shape and properties of quantum black holes from first principles is highly challenging: it requires resumming quantum-gravitational fluctuations, deriving an effective action or a similar mathematical object parametrized by finitely many free parameters, and finally determining the spacetime solutions to the corresponding field equations. In turn, computing the effective action requires solving either the path integral or its integro-differential re-writing in terms of functional renormalization group (FRG)~\cite{Dupuis:2020fhh} equations. To avoid these complications, the so-called renormalization group (RG) improvement has been used extensively in the framework of asymptotically safe gravity to investigate how quantum-gravitational effects could impact the short-distance behavior of gravity beyond GR and its solutions. This approach has emerged in the context of gauge theories~\cite{Coleman1973:rcssb,Migdal:1973si,Adler:1982jr,Dittrich:1985yb} as a way to access leading-order quantum effects while avoiding the computation of quantum loops or a full solution of the beta functions. It consists of promoting the classical constants to running couplings and subsequently replacing the RG scale with a characteristic energy scale of the system. 

At a qualitative level, the application of the RG improvement to gravity~\cite{Bonanno:2006eu,Falls:2010he,Cai:2010zh,Falls:2012nd,torres15,Koch:2015nva,Bonanno:2015fga,Emoto:2005te,Bonanno:2016rpx,Kofinas:2016lcz,Falls:2016wsa,Bonanno:2016dyv,Bonanno:2017gji,Bonanno:2017kta,Bonanno:2017zen,Bonanno:2018gck,Liu:2018hno,Majhi:2018uao,Anagnostopoulos:2018jdq,Adeifeoba:2018ydh,Pawlowski:2018swz,Gubitosi:2018gsl,Bonanno:2019ilz,Held:2019xde,Platania:2019qvo,Platania:2019kyx,Ishibashi:2021kmf,Chen:2022xjk,Scardigli:2022jtt} has pointed to the following tentative conclusions (see~\cite{Platania:2023srt} for a review). First, classical static black holes are replaced by regular black holes~\cite{Bonanno:1998ye,Bonanno:2000ep,Cai:2010zh,Falls:2010he,Torres:2014gta,Kofinas:2015sna,Emoto:2005te,Torres:2017ygl,Adeifeoba:2018ydh, Pawlowski:2018swz, Platania:2019kyx} or by compact objects~\cite{Bonanno:2019ilz,Borissova:2022jqj}. Secondly, accounting for the formation of black holes from the gravitational collapse of a massive star makes singularity resolution less straightforward and typically results in a weaker condition: black hole singularities are not fully resolved, but are rather replaced by so-called integrable singularities~\cite{Fayos:2011zza,torres15,Bonanno:2016dyv,Bonanno:2017kta,Bonanno:2017zen}. Thirdly, singularity resolution in cosmology leads to either bouncing cosmologies or to cyclic universes~\cite{Kofinas:2016lcz, Bonanno:2017gji}. Finally, in a cosmological context, the spectrum of temperature fluctuations in the cosmic microwave background radiation is intuitively understood in terms of fundamental scale invariance~\cite{Wetterich:2019qzx,Wetterich:2020cxq} in the UV---a key requirement for a theory to be UV-complete at a fixed point of the RG flow, cf.~\cite{Bonanno:2001xi,Bonanno:2001hi,Bonanno:2002zb,Guberina:2002wt,Reuter:2005kb,Bonanno:2007wg,Bonanno:2008xp,Bonanno:2010mk,Cai:2011kd,Bonanno:2015fga,Bonanno:2016rpx,Bonanno:2017pkg,Bonanno:2018gck,Gubitosi:2018gsl,Platania:2019qvo,Platania:2020lqb}. 

Yet, the connection of these results with asymptotic safety and the FRG seems vague, as the application of the RG improvement to gravity is subject to ambiguities. In particular, in the context of gravity it is not obvious how to identify the RG scale consistently, as several characteristic physical scales may compete in a given process or phenomenon. This has led to a plethora of applications of the RG improvement in gravity, where the scale is identified based on physical intuition. In addition to this ambiguity, it is not clear whether the RG improvement should be implemented at the level of the action, field equations, or solutions. While these details typically do not affect the qualitative conclusions obtained via the RG improvement (at least when the scale is reasonably motivated and not manifestly inconsistent, e.g., with diffeomorphism invariance~\cite{Babic:2004ev}), a more rigorous approach might allow to determine the connection of these results with first-principle computations in quantum gravity, and in particular with the form factors program~\cite{Knorr:2018kog,Knorr:2019atm,Bosma:2019aiu,Draper:2020knh,Draper:2020bop,Knorr:2021niv,Knorr:2021iwv,Knorr:2022lzn}. The importance of the latter lies in the possibility to compute (via FRG calculations) the effective action in a curvature expansion, including infinitely many higher-derivative terms, and thus to determine formal properties of the theory~\cite{Gies:2016con,Draper:2020bop,Platania:2020knd,Knorr:2021slg,Knorr:2021niv,Bonanno:2021squ,Fehre:2021eob,Platania:2022gtt,Pastor-Gutierrez:2022nki} and of its solutions~\cite{Bosma:2019aiu}.

The scope of this work is to fill the gap between such FRG calculations and the current practise of the RG improvement. We do so by exploiting the so-called decoupling mechanism~\cite{Reuter:2003ca}\footnote{The decoupling of UV modes from IR physics was originally studied in the context of renormalizable field theories and led to the decoupling theorem by Appelquist and Carazzone~\cite{Appelquist:1974tg}. The decoupling theorem famously applies to the Euler-Heisenberg Lagrangian~\cite{Heisenberg:1936nmg} and Fermi’s theory of weak interactions~\cite{Fermi:1934hr}. In systems where spontaneous symmetry breaking or mixing effects occur, the decoupling does not generically take place. An example is chiral perturbation theory~\cite{Gasser:1984gg}.}: if below a certain critical RG scale---dubbed the decoupling scale---there are infrared (IR) scales dominating over the regulator which implements the Wilsonian integration of quantum gravitational fluctuations, then the RG flow freezes out and the scale-dependent effective action at the decoupling scale provides a good approximation to the effective action. In particular, identifying the decoupling scale typically grants access to some higher-derivative interaction terms which were not taken into account in the original truncation. For instance, this is the case in scalar electrodynamics, where the decoupling mechanism allows to derive the logarithmic interaction term in the Coleman-Weinberg effective potential (see~\cite{Reuter:2003ca,Platania:2020lqb} for details).

In this paper we investigate the first application of the decoupling mechanism in gravity. In particular, we will use it to determine qualitative features of the dynamics of black holes beyond GR, from formation to evaporation. As a first attempt in this direction, we will start from the Einstein-Hilbert truncation and use a simple model for the gravitational collapse where the mass function is linear in the advanced time. 
Our key results can be summarized as follows. The dynamics of quantum-corrected black holes is governed by an effective Newton coupling which decreases both in time (down to a certain non-zero value), and along the radial direction. In particular, its radial dependence smoothly interpolates between the observed value at large distances and zero at the would-be singularity. As a consequence, the curvature of the quantum-corrected spacetime is weakened compared to its classical counterpart. Although we started from the Einstein-Hilbert truncation, the effective Newton coupling also features characteristic damped oscillations reminiscent of black hole solutions in higher-derivative gravity with specific non-local form factors: free oscillations in the lapse function are typical of black holes in local quadratic gravity assuming a specific sign of the Weyl-squared coupling~\cite{Bonanno:2013dja,Bonanno:2019rsq}, whereas their damping requires the presence of non-local form factors in the quadratic part of the action~\cite{Zhang:2014bea}. This is an expected outcome of the decoupling mechanism and provides evidence that a careful application of the RG improvement, where the scale is not set by physical intuition, but rather by rigorously exploiting the decoupling condition, might provide important insights into quantum gravity phenomenology~\cite{Addazi:2021xuf}.
Finally, within some approximations, a standard study of the black hole evaporation leads to conclusions in line with the literature~\cite{Bonanno:2006eu}: in the evaporation process, quantum black holes get hotter, and after reaching a maximum temperature, they start cooling down, eventually resulting in a cold black hole remnant.

The present paper is organized as follows. In Sect.~\ref{sect:FRG-RGimp-DecMech} we introduce the FRG and the decoupling mechanism. Next, we show how the decoupling mechanism can be exploited to access some of the higher-derivative terms in the effective action, and thus how to derive corrections to the solutions of GR. We present our setup in Sect.~\ref{sect:setup}, where we also derive the equations governing the dynamics of the quantum-corrected spacetime. We provide numerical and analytical solutions to these equations in Sects.~\ref{sect:collapse}, \ref{sect:Solutionsstatic}, and \ref{sect:evaporation}, where we study the dynamics of quantum-corrected black holes in three distinct regimes: formation, static configuration at the end of a collapse, and evaporation, whereby we assume that the evaporation starts only after the collapse is over. We discuss our results in Sect.~\ref{sect:conclu}.

\section{Functional renormalization group and decoupling mechanism}\label{sect:FRG-RGimp-DecMech}

This section introduces the key novel ingredient in our derivation of the dynamics of black holes beyond GR: the decoupling mechanism~\cite{Reuter:2003ca}. To this end, we shall start by briefly summarizing the FRG, its relation to quantum field theory, and its use in quantum gravity. Next, we shall clarify the difference between the RG scale built into the FRG and the physical running appearing in the effective action and in scattering amplitudes (see also~\cite{Donoghue:2019clr,Bonanno:2020bil}). Finally, we will review the idea behind the decoupling mechanism and we will explain how it can be exploited to extract qualitative information on quantum spacetimes and their dynamics.

\subsection{Effective actions and the functional renormalization group}

Schwarzschild black holes and the Friedmann-Lema\^{i}tre-Robertson-Walker cosmology can be found as solutions to the Einstein field equations
\begin{equation}
    \frac{\delta S_{\text{EH}}}{\delta g_{\mu\nu}}=0\,,
\end{equation}
$S_{\text{EH}}$ being the classical Einstein-Hilbert action. In a quantum theory of gravity these field equations are replaced by their quantum counterpart,
\begin{equation}\label{eq:field-eqs-eff-act}
    \frac{\delta \Gamma_0}{\delta g_{\mu\nu}}=0\,,
\end{equation}
where $\Gamma_0$ is the gravitational quantum effective action. The knowledge of the effective action thus paves the way to the investigation of quantum black holes and quantum cosmology.

Yet, computing the effective action is extremely challenging. One should solve either the gravitational path integral
\begin{equation}\label{eq:path-integral}
    \int \mathcal{D} g_{\mu\nu}\, e^{i\,S_{\text{bare}}[g_{\mu\nu}]} \, ,
\end{equation}
equipped with a suitable regularization, or the FRG equation. Within the FRG, the idea is to first regularize the path integral by introducing an \emph{ad hoc} regulator term in the bare action, and then transform the integral over field configurations~\eqref{eq:path-integral} into a functional integro-differential equation for a scale-dependent version of the effective action, $\Gamma_k$, called effective average action. The resulting flow equation for $\Gamma_k$~\cite{Wetterich:1992yh,Reuter:1996cp} reads
\begin{equation}\label{eq: flow equation}
k \partial_k \Gamma_k = \frac{1}{2} \text{STr}\qty[\qty(\Gamma_k ^{(2)}+\mathcal{R}_k)^{-1}k \partial_k \mathcal{R}_k]\,.
\end{equation}
Here $\Gamma_k ^{(2)}$ denotes the matrix of second functional derivatives of the effective average action with respect to the quantum fields at fixed background. The function $\mathcal{R}_k$ is a regulator whose properties guarantee the suppression of IR and UV modes in the flow equation, such that the main contribution to $\Gamma_k$ comes from momentum modes at the scale $k$. Finally, the supertrace ``STr'' denotes a sum over discrete indices as well as an integral over momenta.

The solution to Eq.~\eqref{eq: flow equation} for a given initial condition identifies a single RG trajectory. The set of all RG trajectories defines the RG flow. A solution $\Gamma_k$ is physically well defined (i.e., the corresponding theory is renormalizable) if its RG trajectory approaches a fixed point in the UV, $k\to\infty$. In this limit~$\Gamma_k$ ought to approach the bare action $S_{\text{bare}}$, up to the reconstruction problem, see, e.g.,~\cite{Manrique:2008zw,Morris:2015oca,Fraaije:2022uhg}. The opposite limit, $k\to0$, corresponds to the case where all quantum fluctuations are integrated out, and yields the standard quantum effective action $\Gamma_0$. First steps towards computing the gravitational effective action have been taken in~\cite{Codello:2015oqa,Knorr:2018kog,Knorr:2019atm,Ohta:2020bsc,Bonanno:2021squ,Basile:2021krr,Knorr:2021niv} in the context of asymptotically safe gravity and in~\cite{Fradkin:1985ys,Gross:1986mw,Veneziano:1991ek,Meissner:1991zj,Meissner:1996sa,Tseytlin:2006ak,Hohm:2015doa,Hohm:2019ccp,Hohm:2019jgu,Basile:2021euh,Basile:2021krk,Hu:2022myf} within string theory. While deriving the coefficients and form factors in the effective action is highly challenging, one may attempt to find solutions to Eq.~\eqref{eq:field-eqs-eff-act} using alternative strategies. Before describing one of them, that is based on the decoupling mechanism, in the next subsection we shall first clarify a fundamental difference between the momentum scale $k$ in Eq.~\eqref{eq: flow equation} and the physical momentum dependence of $\Gamma_k$, as this difference is often a source of confusion.

\subsection{Clarifying nomenclature: RG scale dependence versus physical running}

The effective average action $\Gamma_k$ is constructed as an RG scale dependent action functional, where all couplings or functions are promoted to $k$-dependent quantities. The action $\Gamma_k$ can thus be parametrized by an infinite-dimensional coordinate vector containing the couplings associated with all possible diffeomorphism-invariant operators. In full generality, the flow equation~\eqref{eq: flow equation} can be associated with infinitely many ordinary coupled differential equations for the couplings. However, in practice the technical complexity requires a truncation of the theory space to a manageable subspace. For instance, at quadratic order in a curvature expansion one has
\begin{equation}\label{eq:EAA-quadratic}
    \Gamma_k=\int \dd[4]x \sqrt{-g}\left(\frac{1}{16\pi G_k}\qty(R-2\Lambda_k)+R\, g_{R,k}(\Box)\,R+ C_{\mu\nu\sigma\rho}\,g_{C,k}(\Box)\,C^{\mu\nu\sigma\rho}\right) \,,
\end{equation}
where $G_k=g_k k^{-2}$ and $\Lambda_k=\lambda_k k^{2}$ are the RG scale dependent versions of the Newton and cosmological constants, $g_k$ and $\lambda_k$ being their dimensionless counterparts, and $g_{R,k}(\Box)$ and $g_{C,k}(\Box)$ are quartic couplings which can generally depend on the d'Alembertian operator. The $k$-dependence is attached with the Wilsonian integration of fluctuating modes from the UV to the IR. In particular, it is typically used to study the fixed point structure of the action, as the existence of suitable fixed points relates to  renormalizability and guarantees that observables computed using the effective action $\Gamma_0$ are finite. Provided that such a suitable fixed point exists, one can integrate the flow down to the physical limit $k=0$, where the effective average action reduces to the quantum effective action $\Gamma_0$.

It is important to remark that the $k$-dependence is not related to the physical momentum dependence of couplings, which is to be read off from the effective action $\Gamma_0$. Specifically, the structure of the effective average action~\eqref{eq:EAA-quadratic} should be contrasted with that of the effective action~\cite{Knorr:2019atm}
\begin{equation}\label{eq:eff-action}
    \Gamma_0=\int \dd[4]x \sqrt{-g}\left(\frac{1}{16\pi G_N}\qty(R-2\Lambda)+R\,\mathcal{F}_R(\Box)\,R+ C_{\mu\nu\sigma\rho}\,\mathcal{F}_C(\Box)\,C^{\mu\nu\sigma\rho}\right) \,,
\end{equation}
where the Newton coupling and the cosmological constant are constants whose values are fixed by observations, while the physical running---encoded in the form factors $\mathcal{F}_i(\Box)\equiv g_{i,0}(\Box)$---is attached to the couplings related to the terms at least quadratic in curvature. Note that the dependence on the d'Alembertian is the curved-spacetime generalization of the dependence of couplings on a physical momentum~$p^2$~\cite{Knorr:2019atm}.

The so-called RG improvement was originally devised as a method to obtain an approximation to the effective action~\eqref{eq:eff-action} (or to the solutions to its field equations) by starting from its $k$-dependent counterpart~\eqref{eq:EAA-quadratic} (typically a local version of it) and subsequently replacing the RG scale $k$ with a physical momentum or energy scale. This seems to be a viable strategy in the context of gauge and matter theories~\cite{Coleman1973:rcssb,Migdal:1973si,Adler:1982jr,Dittrich:1985yb}. 
In the framework of quantum field theory the RG improvement originated from the solutions to the Callan-Symanzik equation, while its relation with the FRG is made concrete by the decoupling mechanism, which we review in the following subsection.

\subsection{Effective actions from the decoupling mechanism}

The flow of the effective average action $\Gamma_k$ from the UV fixed point to the physical IR is governed by the FRG equation~\eqref{eq: flow equation}. In particular, the variation of $\Gamma_k$ on the left-hand side of~\eqref{eq: flow equation} is induced by the artificial regulator $\mathcal{R}_k$. The latter is an effective mass-square term, $\mathcal{R}_k\sim k^2$, suppressing fluctuations with momenta $p^2\lesssim k^2$. 

\begin{figure}[t]
\centering 
\includegraphics[width=0.55\textwidth]{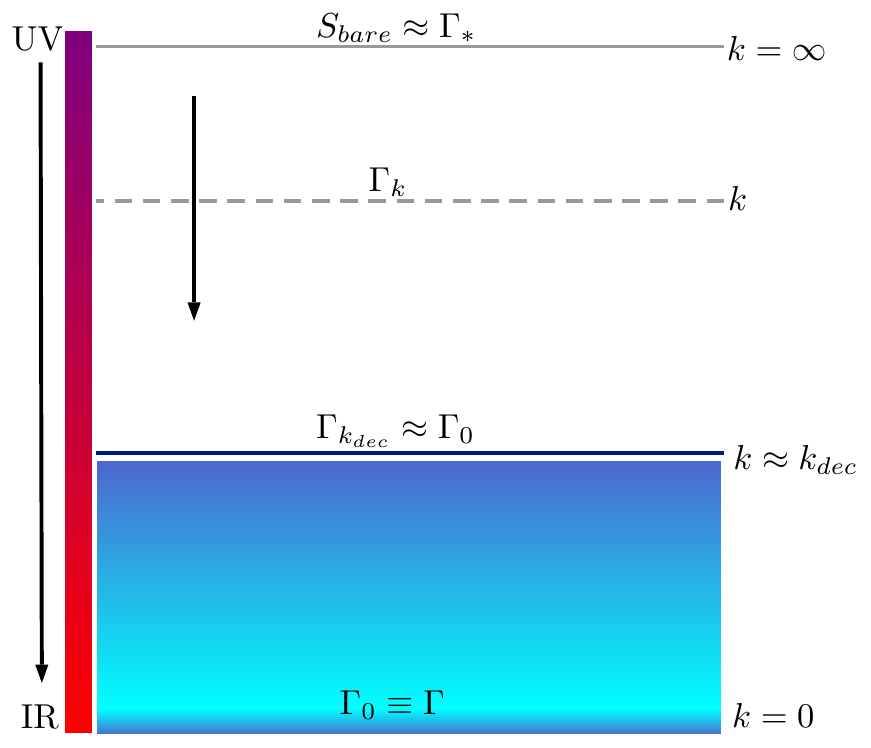}
\caption{\label{fig:decoupl} Idea behind the decoupling mechanism. If one or a combination of physical IR scales in the effective average action overcomes the effect of the regulator $\mathcal{R}_k$ in the flow equation~\eqref{eq: flow equation}, the flow freezes out and the effective average action at the critical scale $k_{dec}$ approximates the full effective action~$\Gamma_0$.}
\end{figure}

The decoupling mechanism~\cite{Reuter:2003ca}, if at work, could provide a short-cut linking (a truncated version of) $\Gamma_k$ to the effective action $\Gamma_0$ and relies on the following observation.
In the flow towards the IR, $\mathcal{R}_k$ decreases as $\sim k^2$, and at a certain scale $k_{dec}$ the running couplings and other physical scales in the action, for instance a mass term, may overcome the effect of the cutoff $\mathcal{R}_k$. As a result, the flow of the effective average action $\Gamma_k$ would freeze out, so that at the decoupling scale $\Gamma_{k=k_{dec}}$ approximates the standard effective action~$\Gamma_0$ (cf.~Fig.~\ref{fig:decoupl}).
By identifying the decoupling scale, certain terms appearing in the full effective action can be predicted which were not contained in the original truncation. An emblematic example is scalar electrodynamics, where the RG improvement, combined with the decoupling mechanism, is able to correctly generate the logarithmic corrections in the Coleman-Weinberg effective potential~\cite{Coleman1973:rcssb}, see~\cite{Reuter:2003ca,Platania:2020lqb} for details and~\cite{Migdal:1973si,Adler:1982jr,Dittrich:1985yb} for other examples in the context of quantum electrodynamics and quantum chromodynamics.

It is important to notice that generally $k_{dec}$ will depend on a non-trivial combination of several physical IR scales appearing in the action, e.g., curvature invariants, masses, or field strengths. To make this statement more precise, one has to look at the structure of the regularized inverse propagator. Schematically, i.e., neglecting any tensorial structure, the latter is conveniently written as
\begin{equation}\label{eq: invProp}
    \Gamma^{(2)}_k+\mathcal{R}_k= c\,( p^2+A_k[\Phi]+\tilde{\mathcal{R}}_k)\,,
\end{equation}
where $c$ is a constant, $\Phi$ denotes the set of fields in the theory defined by $\Gamma_k$, and by definition $A_k[\Phi]\equiv  \Gamma^{(2)}_k/c -p^2 $ and $\mathcal{R}_k\equiv c \, \tilde{\mathcal{R}}_k$. The regulator $\tilde{\mathcal{R}}_k$ efficiently suppresses modes with $p^2\lesssim k^2$ when it is the largest mass scale in the regularized inverse propagator. By contrast, if $A_k[\Phi]$ contains physical IR scales, there might be a critical momentum~$k_{dec}$ below which~$\tilde{\mathcal{R}}_k$ becomes negligible. Grounded on these arguments, the decoupling condition reads
\begin{equation}\label{eq:dec-condition}
    \tilde{\mathcal{R}}_{k_{dec}} \approx A_{k_{dec}}[\Phi] \,,
\end{equation}
and provides an implicit definition of the decoupling scale $k_{dec}$. We emphasize that whereas the structural form of the inverse propagator~\eqref{eq: invProp} defines the functional $A_k$ for general $k$, at the decoupling scale $k_{dec}$, $A_{k_{dec}}$ satisfies the condition~\eqref{eq:dec-condition}.  It is worth noticing that this equation might not have real solutions, in which case the RG improvement would not be applicable.

We finally remark that the implementation of the RG improvement at the level of the Callan-Symanzik equation and in the EAA is subtlety different: in the former the decoupling occurs due to a balance of  physical scales only, while in the latter the unphysical momentum~$k$ is involved. Since the first works independently of the choice of cutoff, the second might inherit this property too.

\subsection{Decoupling mechanism versus the practice of RG improvement}

The procedure of RG improvement consists of promoting the coupling constants to RG scale dependent couplings, and ``identifying'' the IR cutoff $k$ with a physical scale in order to capture qualitatively the effect of higher-order and non-local terms in the full effective action~\eqref{eq:eff-action}.

The scale dependence is governed by the beta functions which can be computed using functional RG methods within a given truncation of the effective action. For instance, if the effective average action for gravity is approximated by monomials up to first order in the curvature, its form reduces to the Einstein-Hilbert action with a scale-dependent Newton coupling and cosmological constant. Neglecting the cosmological constant, the flow equation~\eqref{eq: flow equation} for the running Newton coupling gives rise to the approximate scale dependence~\cite{Bonanno:2000ep}\footnote{Although the RG scale dependence~\eqref{eq: G running} has been first derived in~\cite{Bonanno:2000ep} via computations in Euclidean signature, Eq.~\eqref{eq: G running} still appears to be a good approximation in Lorentzian signature~\cite{Fehre:2021eob}.}
\begin{equation}\label{eq: G running}
 G(k) = \frac{G_0}{1+\omega G_0 k^2},
\end{equation}
where $\omega = 1/g_*$, $g_*$ being the non-Gaussian fixed point value of the dimensionless Newton coupling $g(k) = G(k)k^2$. The existence of such a fixed point in the UV, combined with the requirement of a finite number of relevant directions, are key requirements for the definition of asymptotically safe theories. 

The application of the RG improvement in gravity suffers from the following problems:

\begin{itemize}
    \item \emph{The challenge to relate the  cutoff $k$ to a physical scale of the system}: A priori, the role of $k$ is to provide a way to parametrize the RG flow from the UV towards the IR. In principle, if one ignores the original idea behind the decoupling mechanism, there exists no general prescription of how to perform the scale identification on curved spacetimes and in situations where spacetime symmetries are insufficient to fix $k$ uniquely. Scale-setting procedures relying on diffeomorphism invariance and minimal scale dependence of the action were proposed in~\cite{Reuter:2003ca,Babic:2004ev,Domazet:2010bk,Koch:2010nn,Domazet:2012tw,Koch:2014joa}, but these are either not always applicable, or they provide insufficient information to completely fix the function~$k(x)$. Moreover, in generic physical situations, there is more than one scale.
    \item \emph{Ambiguity in the application at the level of action, field equations, and solutions}: Although the RG improvement in gravity is motivated by the decoupling of the RG flow of the action functional $\Gamma_k$, the sequence of replacements $g_i\to g_i(k)\to g_i[k(x)]$ can in principle be implemented at the level of the action, field equations, or solutions. Typically, the latter two implementations are more straightforward than that at the level of the action, since in physical applications one could skip the step of deriving the field equations or their solutions, respectively. Nevertheless, the three procedures can yield different results. This can be intuitively understood, as, for instance, the replacement $k\mapsto k(x)$ at the level of the action would generate higher-derivative operators which would in turn reflect in additional terms in the field equations.
    \item \emph{Backreaction effects in gravity}: In the context of quantum field theories other than gravity, the RG improvement can be applied straightforwardly~\cite{Coleman1973:rcssb,Migdal:1973si,Adler:1982jr,Dittrich:1985yb}. The reason is that in this case coordinates and momenta are related trivially, $p\sim 1/r$, and the background metric is fixed and typically flat. In the context of gravity this procedure is less controlled, since the definition of any diffeomorphism-invariant (i.e., scalar) quantity requires a metric. Classical spacetimes are however singular, and their metric cannot be trusted in the proximity of the would-be singularities. Moreover, the Newton coupling itself, which in a standard RG improvement procedure is supposed to be replaced with its running counterpart, is part of the metric needed for the definition of the map $k(x)$. Finally, when a one-step RG improvement is performed at the level of the solutions, this induces a change in the effective Einstein equations and a change in the spacetime metric, which in turn would lead to a different map $k(x)$. This suggests that the application of the RG improvement in gravity requires taking backreaction effects into account and determining the effective metric self-consistently, e.g., via the iterative procedure devised in~\cite{Platania:2019kyx}. 
\end{itemize}
Summarizing, the RG improvement in gravity was originally motivated by the decoupling mechanism, and the scale identification $k(x)$ was meant to act as a short-cut to determine the effective action. Yet, in most works on RG improved spacetimes the function $k(x)$ has been fixed based on physical intuition only, accounting neither for the consistency constraints stemming from Bianchi identities (aside from a few examples, e.g.~\cite{Reuter:2003ca,Babic:2004ev,Domazet:2010bk,Koch:2010nn,Domazet:2012tw,Koch:2014joa}), nor for the decoupling mechanism. In this work we will for the first time exploit the decoupling mechanism to derive the function $k(x)$ from the decoupling condition~\eqref{eq:dec-condition} and to determine the dynamics of quantum-corrected black holes.

\section{Modified spacetimes from the decoupling mechanism: setup}\label{sect:setup}

In this section we derive the differential equations describing the evolution of the metric of a spherically-symmetric, asymptotically flat black hole spacetime, including quantum corrections computed by exploiting the gravitational beta functions and the decoupling mechanism.

\subsection{Generalized Vaidya spacetimes} \label{sec: classical Vaidya spacetimes}

One of the key lessons of Einstein's GR is the formation of black holes from the gravitational collapse of matter and radiation. Scenarios for the  collapse of a sufficiently massive object have been developed and are discussed controversially in relation to the cosmic censorship conjecture~\cite{Penrose:1969pc}. In its weak form, the conjecture posits that the maximal Cauchy development possesses a complete future null infinity for generic initial data. In other words, an event horizon should exist which prevents an observer at future null infinity from seeing the  singularity. The conjecture is however known to be violated in various models for the gravitational collapse. In particular, well-known classical models which violate this conjecture are the Tolman-Bondi spacetime for the spherical collapse of dust clouds~\cite{Tolman:1934za,Bondi:1947fta,Eardley:1978tr} or the imploding Vaidya spacetime describing the spherical collapse of radiation~\cite{Vaidya:1951zza,Vaidya1966AnAS}. In the latter case the singularity appears when the ingoing radiation hits a chosen spacetime point---typically the origin of the given coordinate system. In this classical model the singularity is initially naked provided that the rate of concentration of the radiation is sufficiently low~\cite{Kuroda:1984}. In the following we will introduce these Vaidya spacetimes as well as an important generalization of the corresponding class of metrics that will be key in our construction. 

The classical imploding Vaidya solution in advanced Eddington-Finkelstein coordinates reads~\cite{Vaidya:1951zza,Vaidya1966AnAS}
\begin{equation}\label{eq: Vaidya metric classical}
\dd{s^2} = -f(r,v)\dd{v^2} + 2 \dd{v}\dd{r} + r^2\dd{\Omega^2} \,,
\end{equation}
with the lapse function
\begin{equation}\label{eq: lapse function classical}
f(r,v) = 1-\frac{2 G_0 m(v)}{r}\,,
\end{equation}
where $G_0$ denotes the observed value of the Newton coupling. The mass function $m(v)$ depends on the advanced time coordinate and can be used to model a gravitational collapse or evaporation. The metric~\eqref{eq: Vaidya metric classical} with lapse function~\eqref{eq: lapse function classical} is an exact solution to the Einstein equations with vanishing cosmological constant,
\begin{equation}\label{eq: field equations}
G_{\mu\nu} = 8\pi G_0 T_{\mu\nu}\,,
\end{equation}
and an energy-momentum tensor corresponding to a pressureless perfect fluid~\cite{Vaidya:1951zza,Wang:1998qx},
\begin{equation}\label{eq: energy-momentum tensor Vaidya}
T_{\mu\nu} =  \mu \,u_\mu u_\nu \,.
\end{equation}
Here $u^\mu$ is the fluid's four-velocity and
\begin{equation}\label{eq: energy density classical}
\mu = \frac{\dot{m}(v)}{4\pi r^2}
\end{equation}
is its energy density. A dot denotes differentiation with respect to the advanced time. More generally, one could consider a generalized mass function depending both on the advanced time $v$, as well as on the radial coordinate~$r$. The resulting generalized Vaidya spacetime~\cite{Wang:1998qx} is described by a metric of the form~\eqref{eq: Vaidya metric classical} with lapse function
\begin{equation}\label{eq: lapse function}
    f(r,v)=1-\frac{2M(r,v)}{r}\,,
\end{equation}
where the Newton constant $G_0$ is now absorbed in the generalized mass function $M(r,v)$.
These spacetimes are solutions to the classical Einstein equations~\eqref{eq: field equations} with an effective energy momentum tensor
\begin{equation}\label{eq: energy-momentum tensor for generalized Vaidya}
T_{\mu\nu} = \mu \,l_\mu l_\nu + \qty(\rho + p)\qty(l_\mu n_\nu+l_\nu n_\mu) + p g_{\mu\nu}\,,
\end{equation} 
where the two null vectors $l_\mu$ and $n_\mu$ satisfy $l_\mu n^\mu = -1$. The functions $\mu $ and $\rho$ are the two contributions to the energy density associated with the first advanced time and radial derivatives of the generalized mass function $M(r,v)$, while $p$ is the classical pressure computed from its second derivative,
\begin{equation}\label{eq:rhomupclass}
\mu = \frac{\dot{M}(r,v)}{4\pi G_{0} r^2}\,, \quad\quad \rho = \frac{M'(r,v)}{4\pi G_{0} r^2}\,, \quad\quad p = -\frac{M''(r,v)}{8\pi G_{0} r}\,.
\end{equation}
In the previous definition, dots and primes denote derivatives with respect to~$v$ and~$r$, respectively.

Generalized Vaidya spacetimes can be used to model deviations from the Schwarzschild solution. In the next subsection we will make use of these dynamical solutions to describe the collapse of black holes in the presence of quantum-gravitational fluctuations. We will take into account both the backreaction generated by the modifications induced on the spacetime by quantum effects~\cite{Platania:2019kyx}, as well as the dynamical evolution of the spacetime triggered by a time-varying mass function $m(v)$. To this end, we will combine the techniques developed in~\cite{Platania:2019kyx}, which we review below, with the ideas in~\cite{Bonanno:2006eu,Bonanno:2016dyv,Bonanno:2017zen,Bonanno:2017kta}, and with the decoupling mechanism~\cite{Reuter:2003ca,Platania:2020lqb}. 

\subsection{Determining the decoupling scale}

The scope of this subsection is to determine the decoupling scale $k_{dec}$ at which the RG scale dependent effective action $\Gamma_k$ approximates the full effective action. To this end, one first needs to derive the Hessian $\Gamma_k^{(2)}$ and its regularized version.  
Within the Einstein-Hilbert truncation the effective average action is given by 
\begin{equation}\label{eq: effective action}
\Gamma_k =  \int \dd[d]x \sqrt{g}\qty(\frac{1}{16\pi G_k}\qty(2\Lambda_k-R) + \mathcal{L}_m) \,,
\end{equation}
where $G_k$ and $\Lambda_k$ are the Newton coupling and cosmological constant, $d$ is the number of spacetime dimensions, and $\mathcal{L}_m$ is a matter Lagrangian. In our case, since one of our main scopes is to describe the quantum-corrected gravitational collapse, we limit ourselves to the Lagrangian of a perfect fluid with energy-momentum tensor~\eqref{eq: energy-momentum tensor Vaidya}, which reads~\cite{Ray:1972}
\begin{equation}\label{eq: matter Lagrangian}
\mathcal{L}_m = \mu(r,v) \,,
\end{equation}
$\mu(r,v)$ being the energy density of the pressureless fluid as defined in~\eqref{eq: energy density classical}.  The quadratic part of the action is constructed by writing the metric as $g_{\mu\nu}=\bar{g}_{\mu\nu}+h_{\mu\nu}$, where $\bar{g}$ is the background metric and $h$ describes fluctuations about this background, and by expanding the action about $\bar{g}$ up to quadratic order in $h$. The metric fluctuations are split as $h_{\mu\nu}=h_{\mu\nu}^{TL}+d^{-1}\bar{g}_{\mu\nu}\phi$, where $\phi\equiv\bar{g}^{\mu\nu}h_{\mu\nu}$ is the trace part of the metric and $\bar{g}^{\mu\nu}h_{\mu\nu}^{TL}=0$ expresses the orthogonality condition of the traceless mode $h_{\mu\nu}^{TL}$. Further restricting the background to a maximally symmetric spacetime\footnote{This choice is not ideal for a generic black hole background. However, it significantly simplifies the expressions and we do not expect it to impact the qualitative aspects of our results. This expectation comes from two independent considerations. First, within the model of gravitational collapse that we will employ, the spacetime is initially a flat Minkowski background. Thus, a maximally symmetric background is a consistent choice for the early stages of the gravitational collapse. Thereafter, deviations from a maximally symmetric background ought to be automatically encoded in the dynamical adjustment of all physical energy scales and equations involved.
Secondly, we checked that in $d=4$ the corrections induced by a generic Vaidya background would only change the numerical prefactors of our expressions---at least to leading order in the radial coordinate, in the two opposite regions $r\ll l_{Pl}$ and $r\gg l_{Pl}$.} and using a harmonic gauge fixing, with
\begin{equation}
    \Gamma_{gf}=\frac{1}{2} \int \dd[d]x \sqrt{g}\frac{1}{16\pi G_k} \qty[\bar{g}^{\mu\nu}\qty(\bar{D}^\sigma h_{\mu\sigma}-\frac{1}{2}\bar{D}_\mu \bar{g}^{\alpha\beta}{h}_{\alpha\beta})\qty(\bar{D}^\rho h_{\nu\rho}-\frac{1}{2}\bar{D}_\nu \bar{g}^{\alpha\beta}{h}_{\alpha\beta})]\,,
\end{equation}
the regularized Hessian becomes diagonal in field space. Its elements in the trace, traceless, and Faddeev–Popov ghost sectors~\cite{Reuter:2019byg} are
\begin{equation}
\begin{aligned}
    & \left.\Gamma_k^{(2)}+\mathcal{R}_k\right|_{h}={G_k^{-1}}\left( \Box+k^2 r_k(\Box/k^2)-2\lambda_k+C_T \bar{R}+\mu G_k \right)\,, \\
    & \left.\Gamma_k^{(2)}+\mathcal{R}_k\right|_{\phi}=-\frac{d-2}{2d}{G_k^{-1}}\left(\Box+k^2 r_k(\Box/k^2)-2\lambda_k+C_S \bar{R}+\mu G_k \right)\,,\\
    & \left.\Gamma_k^{(2)}+\mathcal{R}_k\right|_{gh}={G_k^{-1}}\left( \Box+k^2 r_k(\Box/k^2)-2\lambda_k+C_V \bar{R} \right)\,,
\end{aligned}
\end{equation}
where $\Box\equiv-\bar{D}^2$ is the d'Alembertian operator built with background covariant derivatives, $\bar{R}$ is the background Ricci scalar, $r_k$ is the dimensionless version of the regulator $\mathcal{R}_k$, and
\begin{equation}
    C_T=\frac{d(d-3)+4}{d(d-1)}\,,\qquad C_S=\frac{d-4}{d}\,,\qquad C_V=-\frac{1}{d}\,.
\end{equation}
As we are interested in asymptotically flat spacetimes, in the following we shall neglect the contribution from the cosmological constant. At this point, the decoupling condition~\eqref{eq:dec-condition} reads
\begin{equation}\label{eq:decoupl-cond}
G_{k_{dec}}^{-1}(\gamma \bar{R} +G_{k_{dec}} \mu-k^2_{dec} r_{k_{dec}})=0 \,,
\end{equation}
where $\gamma\equiv \text{max}\{|C_T|,|C_V|,|C_S|\}$, and $\gamma=2/3$ for $d=4$. In particular, to determine the form of the decoupling scale a simple mass-type regulator $\mathcal{R}_k\simeq k^2$ suffices. This is tantamount to setting $r_k=1$, so that the decoupling condition reads
\begin{equation}\label{eq:dec-cond}
    k_{dec}^2\equiv G_{k_{dec}} \mu +\gamma \bar{R} \,.
\end{equation}
Accounting for the decoupling condition at the level of the action, in combination with the expression~\eqref{eq: G running} for the running Newton coupling~$G_k$, would thus yield an effective action
\begin{equation}
    \Gamma_0\approx \Gamma_{k=k_{dec}}=\int d^4 x \sqrt{g} \left(-\frac{1+\mu\, \omega \,G_0^2}{16\pi G_0}\,\bar{R}-\frac{\gamma \omega (1-\mu \,\omega\, G_0^2)}{16\pi}\,\bar{R}^2 + \mathcal{O}(\mu^2,\bar{R}^3) \right) \,,
\end{equation}
where we have expanded all terms in a curvature expansion and linearized the final expression with respect to the energy density $\mu$. The resulting effective action contains some of the higher-derivative terms in~\eqref{eq:eff-action}, as expected, as well as a non-minimal coupling with matter, encoded in the terms $\mu \bar{R}$ and $\mu \bar{R}^2$.
We thus expect that the dynamical spacetimes stemming from the implementation of the decoupling mechanism at the level of the solutions---which are the focus of our work---will reflect the presence of the higher-derivative operators and of the non-minimal coupling~$\mu R$. As we will see, our findings are consistent with this expectation. Note that this is non-trivial: in past applications of the RG improvement, accounting neither for the backreaction effects of~\cite{Platania:2019kyx} nor for the decoupling condition, the implementation of the replacement $k\mapsto k(x)$ at the level of the action or at the level of the solutions would generally yield different results.

Before proceeding, we briefly comment on the possibility to further constrain the function $k(x)$ by requiring the validity of the Bianchi identities~\cite{Reuter:2003ca,Babic:2004ev,Domazet:2010bk,Koch:2010nn,Domazet:2012tw,Koch:2014joa}.
The simplest starting point is the RG scale dependent version of the Einstein-Hilbert action, Eq.~\eqref{eq: effective action}. If the stress-energy tensor for the matter is separately conserved, the Bianchi identities impose a consistency condition on the function~$k(x)$. The specific form of the modified Bianchi identities relies on whether one makes the replacement $k\mapsto k(x)$ at the level of the action or at the level of the field equations (see~\cite{Platania:2020lqb} for details). If the scale dependence is first introduced at the level of the action, this condition reads~\cite{Reuter:2003ca}
\begin{equation}\label{eq: Bianchi condition}
2 G(k) \Lambda^\prime(k) + G^\prime(k)\qty(R-2\Lambda(k)) = 0 \,,
\end{equation}
where primes denote the differentiation with respect to $k$. The requirement~\eqref{eq: Bianchi condition} expresses the fact that the sum of the effective energy momentum tensor introduced by the coordinate dependence of the Newton coupling and the cosmological constant term should be conserved to guarantee consistency with the covariant conservation of the Einstein tensor. Such a requirement turns out to be redundant in our case, since the effective spacetimes are solutions to field equations of the form~\eqref{eq: field equations} which are found self-consistently.
We thus conclude that in our case the consistency conditions~\cite{Reuter:2003ca,Babic:2004ev,Domazet:2010bk,Koch:2010nn,Domazet:2012tw,Koch:2014joa} are automatically satisfied and therefore do not add additional constraints.

\subsection{Effective dynamics from the decoupling mechanism}\label{sect:effeqs-sol}

The RG improvement in gravity entails working in a truncated version of the effective average action, where the running of the couplings is determined by their beta functions and the RG scale parameter is to be related to a physical energy scale of the system, e.g.~the decoupling scale.  Nevertheless, due to the complexity of the resulting modified field equations\footnote{Already at quadratic order finding full solutions to the field equations requires extended numerical analyses~\cite{Lu:2015cqa,Lu:2015psa,Lu:2015tle,Bonanno:2019rsq}.}, the RG improvement is sometimes (in particular in the context of black-hole physics) implemented at the level of classical equations or solutions. In the following, we will exploit an RG improvement at the level of classical Vaidya solutions.

In this section we derive the equations governing the effective dynamics of a spherically symmetric black hole by combining the decoupling mechanism with the iterative procedure devised in~\cite{Platania:2019kyx}. The latter replaces the standard RG improvement at the level of the solutions with a self-consistent approach accounting for the backreaction effects generated by the introduction of quantum effects on dynamical spacetimes. We will first review this procedure and will subsequently combine it with the decoupling mechanism to derive quantum-corrected spacetimes of the Vaidya type and to study their dynamics.

The starting point is the classical (static) Schwarzschild spacetime with lapse function given by~\eqref{eq: lapse function}, $m(v)\equiv m$ being the mass of the black hole as measured by an observer at infinity. While the exterior Schwarzschild metric is a solution to the vacuum Einstein equations, a non-zero effective energy-momentum tensor $T_{\mu\nu}$ is expected to be present on the right-hand side of the field equations~\eqref{eq: field equations}. The latter can arise in the presence of (quantum) matter, or via quantum-gravitational effects in the form of higher derivatives in the gravitational effective action, cf. Eq.~\eqref{eq:eff-action}. 
Due to these additional terms, the metric of a static spherically-symmetric black hole is expected to be modified with respect to the classical case. 
Assuming that the time and radial components of the metric are inversely related, $g_{rr}=g_{tt}^{-1}$, as is the case for Schwarzschild black holes, the action of quantum effects can be encoded in the radial dependence of an effective Newton coupling $G(r)$. The radial dependence is introduced via the replacement $G_0 \rightarrow G[k(r)]$, and leads to an effective metric of the form
\begin{equation}\label{eq: RG step 1}
f(r) = 1- \frac{2 m G[k(r)]}{r}\,,
\end{equation}
where $k(r)$ is the map between the RG scale $k$ and the radial coordinate $r$, and is initially constructed by means of the classical metric. The spacetime~\eqref{eq: RG step 1} describes an exact solution to the Einstein equations in the presence of a generalized effective  energy-momentum tensor $T_{\mu\nu}^{\text{eff}}$ with energy density $\rho_{\text{eff}} \propto \partial_r G$ and pressure $p_{\text{eff}} \propto \partial_{r}^2G$~\cite{Wang:1998qx,Platania:2019kyx}. This effective energy-momentum tensor has the role of mimicking the higher-derivative terms in the full quantum effective action~\eqref{eq:eff-action}. 

Yet, in a gravitational context the simple replacement $k\to k(x)$ is not expected to yield a good approximation to the effective field equations~\cite{Platania:2019kyx} since: \emph{(i)} the scalar quantity~$k(r)$ (e.g., the proper distance, or a curvature invariant) is necessarily built on the original Schwarzschild metric which fails to give an accurate description of the spacetime in the region of interest, i.e., where quantum gravity effects are important, \emph{(ii)} the new metric~\eqref{eq: RG step 1} is no longer a solution to the vacuum field equations and this backreaction effect might in turn impact the spacetime metric, and \emph{(iii)} a new scale $k(r)$ built with~\eqref{eq: RG step 1} will not match the function $k(r)$ constructed using the Schwarzschild metric.

This points to the conclusion that in gravity backreaction effects induced
by the replacement $k\to k(r)$ have to be taken into account. This can be done iteratively, until a self-consistent solution is reached. In other words, one should iteratively apply the RG improvement until the scale $k_n(r)$ used to define the lapse function $f_{n+1}(r)$ matches the decoupling scale $k_{n+1}$ constructed using the metric $g_{\mu\nu}^{(n+1)}$ at the step $n+1$.
The iterative procedure is implemented by defining the lapse function $f_n(r)$ at a step $n>0$ as
\begin{equation}
f_{n}(r) = 1- \frac{2 m G_{n}[k_{n-1}(r)]}{r} \, ,
\end{equation}
i.e., in terms of a scale $k_{n-1}(r)$ constructed by means of the metric $g_{\mu\nu}^{(n-1)}$ at the step $n-1$. In general, this will be a function of the first and second derivatives of the effective Newton coupling $G_{n-1}(r)$. If the sequence $\{G_n\}$ defined in this way converges, the fixed function $G_\infty(r)$ satisfies a differential equation which is fully determined by the functional form of the scale $k_n(r)$. Specifically, based on the RG running~\eqref{eq: G running}, this yields the differential equation
\begin{equation}\label{eq: G infinity}
 G_\infty(r) = \frac{G_0}{1+\omega\, G_0 k_\infty ^2(r)}\,,
\end{equation}    
with $k_\infty ^2$ depending on $G_\infty$ and its derivatives. 
In~\cite{Platania:2019kyx} the scale has been fixed to be $k^2 \propto \rho$, giving rise to an analytically solvable first-order ordinary differential equation for $G_\infty$. Its solution is given by
\begin{equation}
G_\infty(r) = G_0\qty(1- e^{-\frac{r^3}{l^3}})\,,
\end{equation}
where $l$ is a length scale of the order of the Planck length $l_P$. As a key result, the limit of the sequence of metrics is described by a Dymnikova black hole~\cite{Dymnikova:1992ux} with a regular de Sitter core.

We now proceed by generalizing the framework in~\cite{Platania:2019kyx} to Vaidya spacetimes~\eqref{eq: lapse function}, including a general dynamical mass $m(v)$ in the lapse function~\eqref{eq: lapse function}. This has the ultimate goal to describe the dynamics of quantum black holes from formation to evaporation. In this, the scale $k(x)$ will be derived by an explicit use of the decoupling mechanism, as this is key to connect the RG scale dependent description~\eqref{eq:EAA-quadratic} with the physics of the effective action~\eqref{eq:eff-action}. Specifically, $k(r)$ should be equated to the decoupling scale $k_{dec}$ in Eq.~\eqref{eq:dec-cond} in order for the metric~\eqref{eq: RG step 1} to be an approximate solution to the field equations stemming from an effective action of the type~\eqref{eq:eff-action}. 
In the case of dynamical spacetimes, one can set up the iterative procedure by using the lapse function
\begin{equation}\label{eq: lapse functions RG-iterated}
f_{n}(r,v) = 1-\frac{2 M_{n}(r,v)}{r} \, .
\end{equation}
Here the generalized mass function $M_{n}(r,v)=m(v)\,G_{n}[k_{n-1}(r,v)]$ is defined by the classical mass function and the running Newton coupling at the step $n$ of the iteration. The metric defined by the lapse function~\eqref{eq: lapse functions RG-iterated} belongs to the class of generalized Vaidya spacetimes~\cite{Wang:1998qx} introduced in Sect.~\ref{sec: classical Vaidya spacetimes}. The corresponding metric satisfies the effective Einstein equations 
\begin{equation}\label{eq: RG-improved field equations}
G_{\mu\nu} ^{n} = 8 \pi G_n T_{\mu\nu} ^{n}\,,
\end{equation}
where the effective energy-momentum tensor takes the form~\eqref{eq: energy-momentum tensor for generalized Vaidya} with  energy densities and pressure redefined as
\begin{equation}\label{eq: energy densities and pressure}
\mu_{n} = \frac{\dot{M}_{n}(r,v)}{4\pi G_{n}(r,v) r^2}\,, \quad\quad \rho_{n} = \frac{M' _{n}(r,v)}{4\pi G_{n}(r,v) r^2}\,, \quad\quad p_{n} = -\frac{M'' _{n}(r,v)}{8\pi G_{n}(r,v) r}\,.
\end{equation}
The effective Newton coupling $G_{n}$ will itself depend on the self-adjusting cutoff $k_{n}$ which needs to be determined by the properties of the spacetime at the previous step of the iteration. In particular, the effective Newton coupling will generally depend on both the radial coordinate~$r$ and the advanced time $v$, $G_n=G_n(r,v)$; in the remainder of this section we will omit this dependence for shortness. Finally, in order to make contact with the FRG and determine solutions which approximate those stemming from the full effective action $\Gamma_0$, we shall fix~$k$ to be the decoupling scale $k_{dec}$. In particular, for a fully consistent implementation of the decoupling mechanism, the decoupling scale has to be built using the iterative procedure detailed above.

Setting $r_k=1$ as before and evaluating the decoupling condition~\eqref{eq:dec-cond} on-shell finally yields the definition of the decoupling scale at the step $n+1$,
\begin{equation}\label{eq: cutoff identification}
k_{n+1} ^2 = G_n\qty(\mu_n + \gamma 16 \pi (\rho_n - p_n))\,,
\end{equation}
where we have dropped the label ``dec'' from the decoupling scale and we have written the background Ricci scalar (for metrics of type~\eqref{eq: lapse functions RG-iterated}) in terms of the generalized energy density and pressure~\eqref{eq: energy densities and pressure}, according to
\begin{equation}\label{eq: Ricci scalar}
R = 4 \frac{M' _{n} (r,v)}{r^2} + 2 \frac{M'' _{n} (r,v)}{r} = 16 \pi G_{n}( \rho_{n} - p_{n})\,.
\end{equation}
As will be important later, we note at this point that the high-energy regime $k\gg m_{Pl}$, where the flow is close to the UV fixed point $g_\ast$ of the dimensionless Newton coupling, corresponds to the large-curvature regime. For a spherically symmetric black hole spacetime this means that the UV fixed point regime corresponds to the region close to the classical singularity, while the IR corresponds to large radii.

Finally, taking the limit $n\to \infty$, the dynamical equation for the effective gravitational coupling becomes %
\begin{equation}\label{eq: G infinity differential equation}
G_\infty = \frac{G_0}{1+ G_0 \omega G_\infty \qty(\mu_\infty  + \frac{2}{3}  16 \pi \qty(\rho_\infty - p_\infty))}\,,
\end{equation}
where $\mu_\infty$, $\rho_\infty$ and $p_\infty$ are defined by~\eqref{eq: energy densities and pressure}, with $M_\infty(r,v) \equiv G_\infty(r,v)\, m(v)$, i.e.,
\begin{equation}\label{eq:explicitdependence}
    \mu_{\infty} = \frac{\dot{m}(v)}{4\pi G_{\infty} r^2}+\frac{m(v)\dot{G}_{\infty}}{4\pi G_{\infty} r^2}\,, \quad\quad \rho_{\infty} = \frac{m(v)\,G' _{\infty}}{4\pi G_{\infty} r^2}\,, \quad\quad p_{\infty} = -\frac{m(v)\,G'' _{\infty}}{8\pi G_{\infty} r}\,.
\end{equation}
Inserting these expressions for the energy densities and pressure in terms of the effective Newton coupling and mass~$m(v)$ in Eq.~\eqref{eq: G infinity differential equation}, we obtain the following second-order non-linear partial differential equation for $G_\infty(r,v)$
\begin{equation}\label{eq: G infinity differential equation rough}
\qty(G_0 \omega m \qty(16\pi r G_\infty '' + 32 \pi G_\infty ' + 3  \dot{G}_\infty)  + 3 G_0 \omega \dot{m}G_\infty + 12\pi  r^2)G_\infty - 12 G_0 \pi r^2 = 0\,,
\end{equation}
where the classical Vaidya mass function $m(v)$ is still to be specified. The partial differential equation~\eqref{eq: G infinity differential equation rough} is our first main result and will be used to determine the dynamics underlying the quantum-corrected gravitational collapse and black hole evaporation. Specifically, once a solution to Eq.~\eqref{eq: G infinity differential equation rough} is found, the resulting spacetime metric takes the form of a generalized Vaidya spacetime with lapse function
\begin{equation}
    f_\infty(r,v)=1-\frac{2\,m(v)\,G_{\infty}(r,v)}{r}\,.
\end{equation}
We will use the framework introduced in this section to study the effective metric in different regimes, from formation to evaporation.

\section{Dynamics of the collapse process}\label{sect:collapse}

As a result of the decoupling mechanism, the dynamics of the effective Newton coupling~$G(r,v)$ is governed by Eq.~\eqref{eq: G infinity differential equation rough} where it remains to specify the Vaidya mass function~$m(v)$. In this work we will use one of the simplest models for the gravitational collapse of a massive star, known as Vaidya-Kuroda-Papapetrou (VKP) model~\cite{Vaidya1966AnAS,Kuroda:1984,Papapetrou:1985}. The same model was considered in~\cite{Bonanno:2016dyv,Bonanno:2017kta,Bonanno:2017zen} to study the quantum-corrected collapse based on a one-step RG improvement not accounting for the decoupling mechanism.
We will present two distinct analytical results showing the expected functional dependence of the effective Newton coupling at early times and for small values of the radial coordinate. Finally, by using these solutions as boundary conditions, together with the requirement of matching the observed value of the Newton constant at large distances and early times, we will provide a complete numerical solution to the partial differential equation~\eqref{eq: G infinity differential equation rough}. Non-trivial corrections to the classical black hole spacetime, which describe the outcome of the gravitational collapse, will be the subject of Sect.~\ref{sect:Solutionsstatic}, while the evaporation will be described separately in Sect.~\ref{sect:evaporation}.

\subsection{Vaidya-Kuroda-Papapetrou collapse model}\label{subsec: VKP spacetime}

\begin{figure}[t]
\centering 
\includegraphics[width=0.55\textwidth]{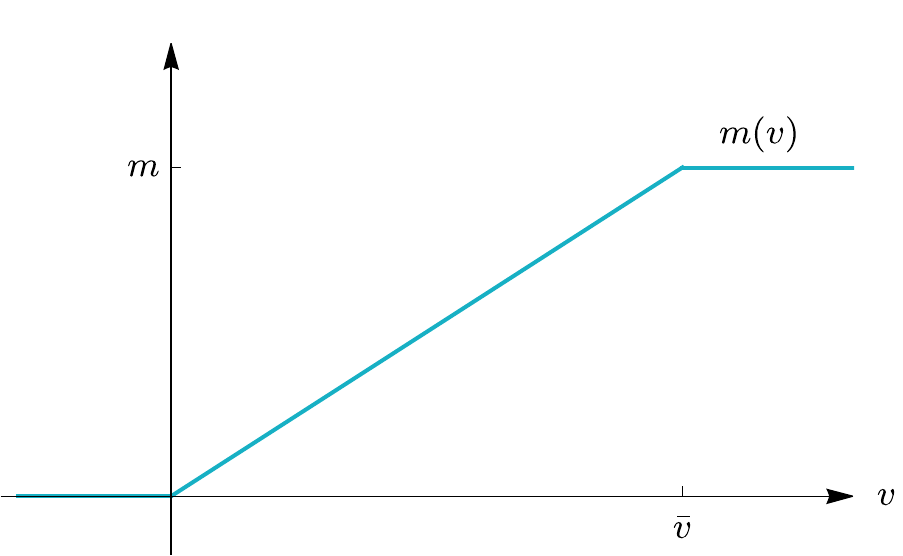}
\caption{\label{fig: mass function VKP} Mass function of the classical VKP spacetime as given in Eq.~\eqref{eq: mass function}. The spacetime is initially flat. At $v=0$ the gravitational collapse starts and the mass $m(v)$ increases linearly with an injection rate $\lambda$. The collapse lasts until $v=\bar{v}$, where the mass function $m(v)$ reaches the plateau $m(v)=m$, $m$ denoting the final mass of the black hole.}
\end{figure}

The VKP spacetime~\cite{Vaidya1966AnAS,Kuroda:1984,Papapetrou:1985} is a simplified model for the gravitational collapse of a massive star. Its geometry is characterized by a linear mass function,
\begin{equation}\label{eq: mass function}
m(v) = \begin{cases}
0, & v \leq 0\,; \\
\lambda v, & 0 < v < \overline{v}\,; \\
m,  & v\geq \overline{v}\,,
\end{cases}
\end{equation}
as shown in Fig.~\ref{fig: mass function VKP}. While for advanced times $v \leq 0$ the spacetime is a flat Minkowski vacuum, at $v=0$ shells of ingoing radiation originating from the star are infused and concentrated towards the origin, $r=0$. The linear increase in mass at the rate $\lambda$ stops at $v=\overline{v}$, when the object settles down to the static classical Schwarzschild spacetime with mass $m$. Historically, the VKP model was one of the first counterexamples to the cosmic censorship conjecture~\cite{Kuroda:1984}.

\subsection{Identifying possible boundary conditions}

In this section we determine the solutions to the dynamical equation~\eqref{eq: G infinity differential equation rough} in two asymptotic regimes, where this equation can be solved analytically. This will provide us with the boundary conditions to solve the full dynamics numerically.

\subsubsection{Dynamics at early times}\label{subsec:ApprVdependence}

The radial dependence of the effective Newton coupling at early times $v\ll \bar{v}$ is dictated by the differential equation~\eqref{eq: G infinity differential equation rough} with the dominant contribution stemming from the energy density associated with $\mu_\infty$. Indeed, for $v\ll \bar{v}$ the spacetime is approximately Minkowski and thus the radial derivatives of the Newton coupling---defining $\rho_\infty$ and $p_\infty$---are approximately zero. Moreover, since during the collapse $m(v)$ is modeled as a power of the advanced time, $m(v)\sim v^n$ (with exponent $n=1$ in our case), it further suppresses $\rho_\infty$ and $p_\infty$, cf. Eq.~\eqref{eq:explicitdependence}. In contrast, $m(v)$ enters $\mu_\infty$ via its advanced time derivative, and since in our case it is a constant, $\dot{m}=\lambda$, its contribution to $\mu_\infty$ will dominate over all terms in $\rho_\infty$ and~$p_\infty$. As a consequence, at early times the effective Newton coupling has only a very weak dependence on the advanced time~$v$, which even drops out if $\rho_\infty$ and~$p_\infty$ are neglected. 

Dropping the $\rho_\infty$ and $p_\infty$ contributions from Eq.~\eqref{eq: G infinity differential equation} the effective Newton coupling reduces to a function of the radial coordinate only, $G_\infty = G_\infty(r)$, and obeys the equation
\begin{equation}\label{eq: G infinity differential equation - approximate}
\qty(3 G_0 \,\omega \, \lambda\,G_\infty(r) + 12\pi r^2)G_\infty(r) - 12 G_0 \pi r^2 = 0\,.
\end{equation}
The positive semi-definite solution to the previous quadratic equation reads
\begin{equation}
G_\infty(r) = \frac{2}{G_0 \lambda \omega}\qty(-\pi r^2 + \sqrt{\pi^2 r^4 + {G_0}^2 \lambda \, \omega \pi r^2})\,. 
\end{equation}
Therefrom, the resulting metric can be determined by inserting the result into the lapse function~\eqref{eq: lapse function}. At small radii, the corresponding Kretschmann scalar scales as
\begin{equation}
R_{\mu\nu\rho\sigma} R^{\mu\nu\rho\sigma} \propto \frac{1}{r^4}\,,
\end{equation}
Compared to the classical curvature singularity $\propto r^{-6}$, the antiscreening of gravity stemming from the fixed point implies a weakening of the curvature singularity already at early times. Moreover, following our later analysis in Sect.~\ref{subsubsec: Solutions static fixed-point}, the divergence of the local curvature at the origin is expected to be further weakened during the collapse. In fact, the curvature for the static solutions at the end of the collapse settles down to a scaling $\propto r^{-3}$.

\subsubsection{Dynamics close to the classical singularity}\label{subsubsec:FPapprox}

We shall in the following consider solutions to the field equations in the region of spacetime close to the would-be singularity, i.e., for $r\ll l_{Pl}$. Our goal is to determine a boundary condition of the form $G_{\infty}(r_{min},v)=J(v)$, at a fixed $r_{min}\ll l_{Pl}$, for the numerical integration of the partial differential equation~\eqref{eq: G infinity differential equation rough}. In contrast to the case~$v\sim0$ studied in the previous subsection, the asymptotic analysis of Eq.~\eqref{eq: G infinity differential equation rough} for $r\sim0$ is extremely involved, and standard techniques based, e.g., on expansions in power laws, are not effective in this case. Yet, as we are only interested in finding a boundary condition $G_{\infty}(r_{min},v)=J(v)$ at a fixed $r_{min}\ll l_{Pl}$, we will utilize two complementary strategies, in combination with some arguments, which we describe in the following.

In our first approach, we neglect the two terms proportional to $r^2$ in Eq.~\eqref{eq: G infinity differential equation rough}, as they are small for $r\sim0$, and dropping them significantly reduces the complexity of the equation. Separating the variables, the ansatz
\begin{equation}\label{eq: G infinity small r ansatz}
G_\infty(r,v) = G_0 \qty(1-H(v))F(r)    
\end{equation}
further simplifies the remaining differential equation to
\begin{equation}\label{eq: F and H together}
3 F(r)(-1+H(v)+v H'(v)) + 16 \pi v (-1 + H(v))(2 F'(r)+r F''(r))) = 0\,,
\end{equation}
which can be rewritten as
\begin{equation}
\frac{2 F'(r) + rF''(r)}{F(r)} = -\frac{3}{16 \pi v}\frac{1-H(v)-vH'(v)}{1-H(v)} \equiv c_0\,.
\end{equation}
Here we have used the fact that the left-hand and right-hand sides of the equation can depend only on $r$ and $v$, respectively, and thus must be equal to a constant $c_0$. As a result we obtain two differential equations determining the functions $F$ and $H$,
\begin{equation}\label{eq: F and H}
\begin{aligned}
2 F'(r) + r F''(r) & =  c_0 F(r)\, ,\\
1- v\frac{H'(v)}{1-H(v)} & = -\frac{16 \pi}{3}c_0 v\, .
\end{aligned}
\end{equation}
The solution for the radial function $F(r)$ is
\begin{equation}\label{eq:r-dependence-small-r}
    F(r)=c_1+\frac{c_2}{r} \,.
\end{equation}
for $c_0=0$, while for $c_0\neq0$ it is given by  
\begin{equation}
   F(r)= c_1 \frac{I_1(2\sqrt{c_0}\sqrt{r})}{\sqrt{c_0}\sqrt{r}}+c_2 \frac{K_1(2\sqrt{c_0}\sqrt{r})}{\sqrt{c_0}\sqrt{r}}\,,
\end{equation}
where $I_n(x)$ and $K_n(x)$ are modified Bessel functions of the first and second kind, respectively, and $c_0>0$ in order for the solution to be real. In both solutions for $F(r)$, $c_1$ and $c_2$ are integration constants. The dependence on the advanced time is instead encoded in the function
\begin{equation}\label{eq: H(v)}
H(v) = 1 + b_1 \frac{e^{-\frac{16\pi}{3}c_0 v}}{v}\, ,
\end{equation}
where $b_1\equiv -v_0$ is an integration constant.
Expanding the exponential function produces a term $\simeq 1/v$ to first order. At next order, $H(v)$ receives a constant contribution. In general, terms coming with an even power in the series expansion of the exponential give rise to positive odd powers of $v$ with a positive pre-factor in the overall expression for~$H(v)$. They would therefore yield positive contributions to the $v$-dependence of the effective Newton coupling and dominate at late times. As we expect the effective Newton coupling to be dynamically weakened during the collapse process, we set these terms to zero by the choice $c_0\equiv 0$ in~\eqref{eq: F and H}, i.e., we proceed by requiring
both summands in Eq.~\eqref{eq: F and H together} to vanish simultaneously. In this case the time dependence is encoded in Eq.~\eqref{eq: H(v)}, with $c_0=0$, while the function $F(r)$ is given by Eq.~\eqref{eq:r-dependence-small-r}, where the integration constant~$c_2$ must be zero, as otherwise the magnitude of the effective Newton coupling would increase towards $r\to0$. This requirement follows from the existence of a fixed point of the RG flow, as in this case the RG scale dependent Newton coupling scales as $G_k\sim g_\ast k^{-2}$ in the UV, and vanishes in the high-energy limit. The anti-screening of the gravitational interaction, making gravity weaker in the UV, is the reason behind the expectation of singularity resolution in asymptotically safe gravity. Finally, the integration constant $c_2$ for the function $F(r)$ is fixed to $c_1 = 1$, which guarantees that the ansatz~\eqref{eq: G infinity small r ansatz} is compatible with the observed value of the Newton constant at early times. In summary, for small radii $r$ and times $v>0$ the form of the effective Newton coupling can be approximated by
\begin{equation}\label{eq: dynamics in UV FP regime}
G_\infty(r,v) = G_0 \frac{v_0}{v}\,.
\end{equation}
This result indicates that the injection of radiation into an initially flat Minkowski spacetime, within the quantum-corrected VKP model, describes a highly non-perturbative process at early times and close to the would-be singularity, after which the strength of the coupling rapidly decreases with time. 

A similar conclusion is also reached by employing an alternative strategy. We shall use once again the ansatz~\eqref{eq: G infinity small r ansatz}, and then proceed by replacing it in the full partial differential equation~\eqref{eq: G infinity differential equation rough} (including the last two terms proportional to $r^2$) and by expanding about $r=0$ up to linear order in the radial coordinate. This procedure yields a differential equation for the function $H(v)$, whose solution reads
\begin{equation}\label{eq:sol-uv-fp-alternative}
    H(v)=1+{b_1}\frac{e^{-\frac{16\pi}{3}c_0 v}}{v}\,,
\end{equation}
where we have defined
\begin{equation}\label{eq:expre-c0}
    c_0=\frac{\left(r\,F(0) F''(0)+2 \,r\, F'(0)^2+2 F(0) F'(0)\right)}{3 F(0) \left(2\, r\, F'(0)+F(0)\right)}\,.
\end{equation}
On the one hand, Eq.~\eqref{eq:sol-uv-fp-alternative} resembles the solution in Eq.~\eqref{eq: H(v)}. On the other hand, its explicit dependence on the radial coordinate $r$---encoded in the expression~\eqref{eq:expre-c0} of $c_0$---shows that a simple separation of variables of the form~\eqref{eq: G infinity small r ansatz}, while generally successful to study the asymptotics of  differential equations, is not effective in our case and leads to contradictions. After all, as already mentioned, the exponent $\alpha$ governing the leading-order scaling of $G_\infty \sim r^\alpha$ for $r\sim 0$ is expected to be a function of the advanced time $v$. The asymptotics~\eqref{eq: dynamics in UV FP regime} is thus not expected to be accurate and the absence of an $r$ dependence in Eq.~\eqref{eq: dynamics in UV FP regime} should not come as a surprise. Specifically, a weak dependence on the radial coordinate at small radii, making the effective Newton coupling vanishing at $r=0$, is expected on physical grounds.

Despite these issues, as the two derivations presented above yield the same $v$ dependence at a fixed spatial slice, and since the aim of this subsection is solely to find a second, reasonable input for the numerical integration, we will assume that Eq.~\eqref{eq: dynamics in UV FP regime} provides a consistent boundary condition at $r=r_{min}\ll l_{Pl}$, and we will use it as an input for the numerical integration. Whether this assumption is consistent can then be verified a posteriori, based on the outcome of the numerical integration. In particular, the $r$ dependence ought to be restored in the full solution. We anticipate here that the numerical solution will be compatible with this expectation, and specifically with an effective Newton coupling that vanishes in the limit $r\to 0$.

\subsection{Full numerical solution} \label{subsubsec:fullsol}

In this section we combine the previous results and provide a numerical solution to the partial differential equation~\eqref{eq: G infinity differential equation rough} for the VKP model. The numerical integration will be performed in the region $(v,r) \in \qty[v_0, \overline{v}] \cross \qty[r_{min},r_{max}]$. Hereby $v_0\ll1$ and $\overline{v}$ denote the start and end time, respectively, for the numerical integration of the equations along the advanced time direction. Similarly, $r_{min}$ and $r_{max}$ are the integration boundaries for the radial coordinate. In particular, for the numerical integration we fixed $v_0/t_{Pl}=0.01$, $\bar{v}/t_{Pl}=1$, $r_{min}/l_{Pl}=10^{-4}$, and $r_{max}/l_{Pl}=50$. Moreover, we set the parameters~$\lambda$ and~$\omega$ to one and we chose the mass $m$ of the black hole to be Planckian, $m/m_{Pl}=1$. In general, in the collapse model introduced in Sect.~\ref{subsec: VKP spacetime}, the infusion rate $\lambda$ and the duration of the collapse~$\bar{v}$ determine the mass~$m$ of the configuration at the end of the collapse. Consequently, different choices of $m$ will have an impact on the properties of the final static object, as we shall see in Sect.~\ref{sect:Solutionsstatic}.

\begin{figure}[t]
\centering
\includegraphics[width=0.5\textwidth]{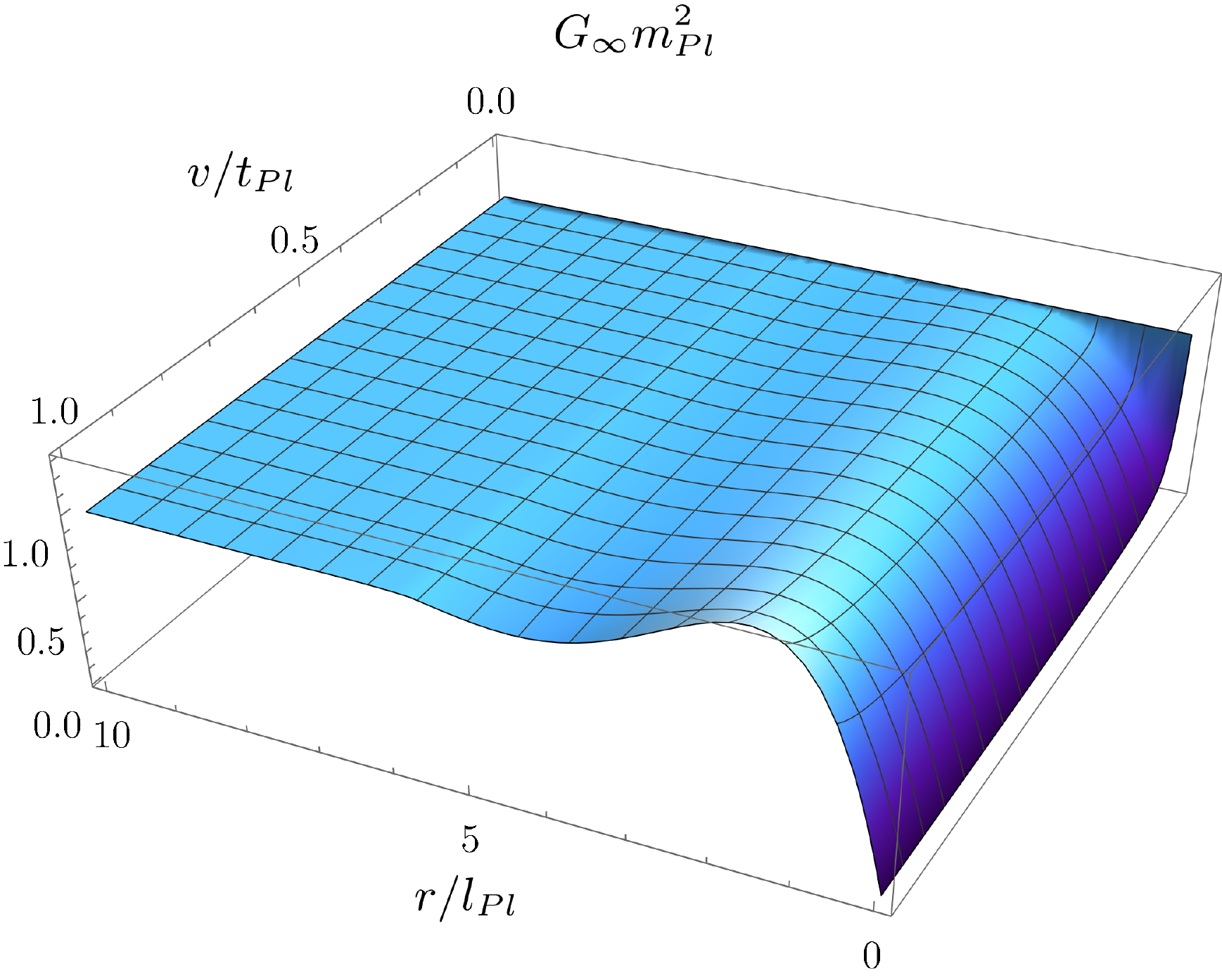}
\caption{\label{fig: numerical solution G infinity}Numerical solution to the partial differential equation~\eqref{eq: G infinity differential equation} for the effective Newton coupling $G_\infty$. We use as initial condition at early times and boundary condition in the IR the observed value of the Newton constant, i.e.~$G_\infty(r,v_0) =  G_\infty(r_{max},v) = G_0$. For the remaining boundary condition near the origin at $r_{min}$ we use the result~\eqref{eq: dynamics in UV FP regime} associated with the dynamics in the proximity of the would-be singularity. At early times for all $r$, as well as at large distances for all $v$, the effective Newton coupling reproduces the observed value of Newton's constant, $G_0$. When the collapse process starts, the effective Newton coupling decays as $\simeq v^{-1}$ until the shell-focusing is over. At the end of the collapse, the effective Newton coupling converges to a function which  interpolates between $G_0$ (large-distance limit) and zero (for small radii). This function features in addition damped oscillations along the radial direction.
}
\end{figure}

In Eq.~\eqref{eq: G infinity differential equation rough} the time derivative and second spatial derivative occur with the same sign on the left-hand side of the partial differential equation, resulting in a structure reminiscent of negative diffusion. It is well known that the numerical analysis of this type of differential equations is very involved. We present here a numerical solution stemming from the following initial and boundary conditions. First, we require that the effective Newton coupling reduces to the observed value $G_0$ both at early times and at large distances. This results in the initial and boundary conditions $G_\infty(r,v_0) =  G_\infty(r_{max},v) = G_0$. Secondly, we make use of the result~\eqref{eq: dynamics in UV FP regime}, which describes the dynamics in the proximity of the classical singularity, to fix the remaining boundary condition near the origin, at $r_{min}/l_{Pl} = 10^{-4}$. More explicitly, this boundary condition reads $G(r_{min},v) = G_0 v_0/v$. Finally, according to our simplified model for the gravitational collapse, see Sect.~\ref{subsec: VKP spacetime}, we evolve the system until a final time $\overline{v}/t_{Pl} = 1$. Fig.~\ref{fig: numerical solution G infinity} shows the result of the numerical integration.

Whereas at early times and large distances the eﬀective Newton coupling is well approximated by its classical value, the situation is drastically diﬀerent for small radii. As soon as the collapse process has started, the effective Newton coupling at
small distances from the radial center becomes weaker, thus providing a direct illustration
of the antiscreening effect of gravity in the UV. Its dependence on the advanced time~$v$ approximately follows an inverse power, cf. Eq.~\eqref{eq: H(v)}. In particular, at the end of the gravitational collapse, the effective Newton coupling interpolates between the classical observed value
$G_0$, which is recovered at large distances, and zero in the limit $r\to0$, i.e., where quantum gravity effects
become stronger. Importantly, we checked that these qualitative features are insensitive to the initial and boundary conditions.
An additional striking feature of the effective Newton coupling lies in its damped oscillations along the radial direction at late times. Such oscillations have been observed in some specific black hole solutions of higher-derivative gravity with specific non-local form factors~\cite{Zhang:2014bea}. This seems to be consistent with the arguments in Sect.~\ref{sect:FRG-RGimp-DecMech}, and specifically with the insight that the decoupling mechanism might  fulfil the original scope of RG improvement, granting access to some of the quantum corrections in the effective action. We will come back to this topic in the next section. 

\begin{figure}[t]
\centering
\includegraphics[width=0.45\textwidth]{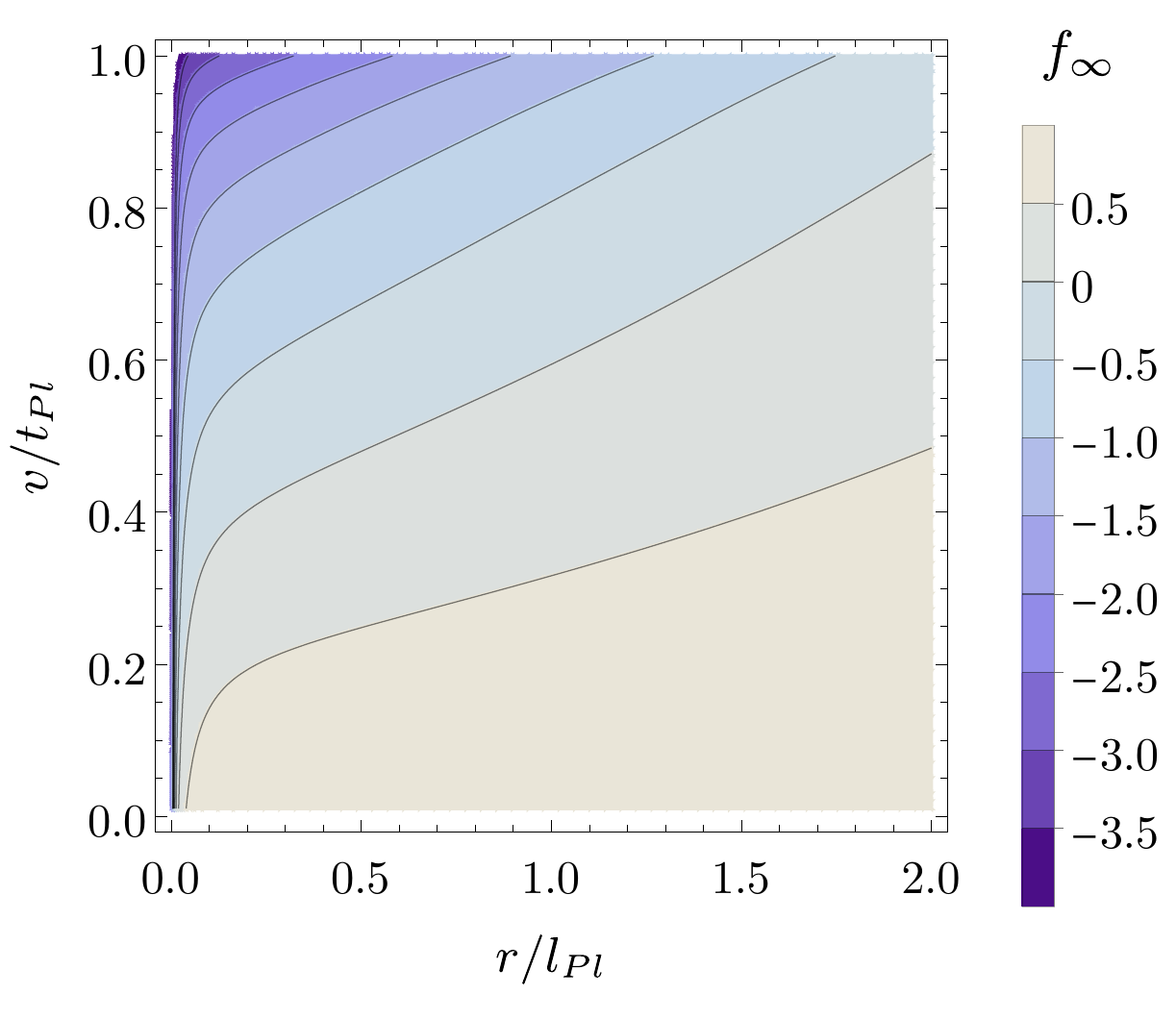}
\hfill
\includegraphics[width=0.53\textwidth]{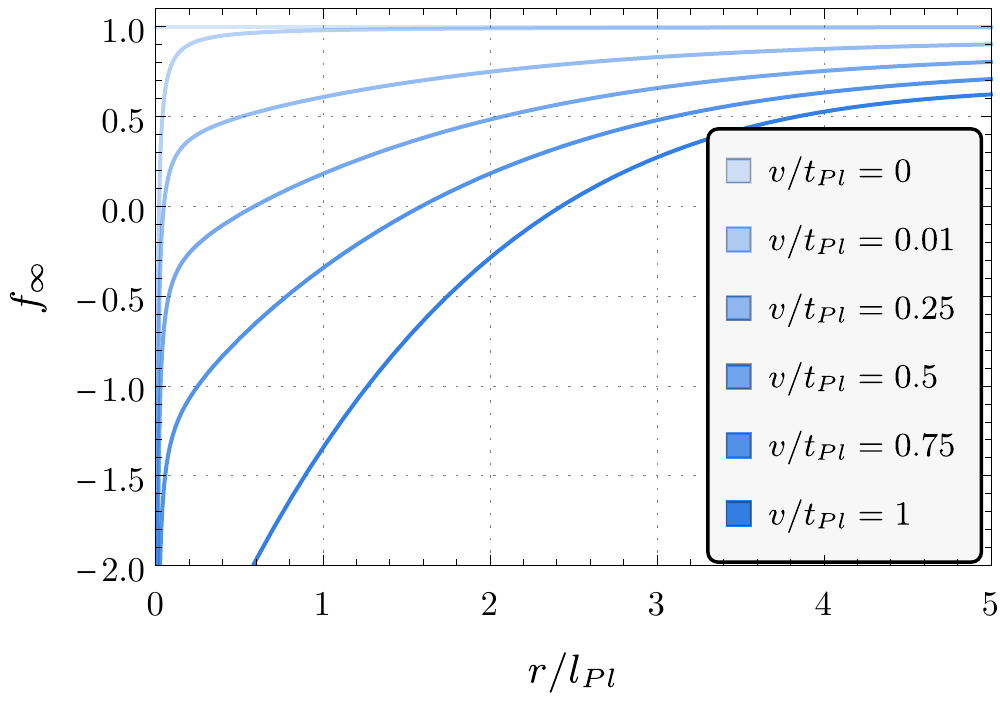}
\caption{\label{fig: numerical solution lapse function}Time evolution and radial dependence of the lapse function $f_\infty(r,v)$, i.e.~the $(0,0)$-component, of the VKP spacetime with effective Newton coupling $G_\infty(r,v)$. The collapse drives the formation of a black hole horizon whose location lies initially at a radius smaller than for the classical VKP model. For sufficiently large masses, the final configuration is approximated well by the Schwarzschild spacetime, at least outside of the Planckian region, i.e., for~$r\gg l_{Pl}$. 
}
\end{figure}

Finally, Fig.~\ref{fig: numerical solution lapse function} shows the time-evolution of the $(0,0)$-component of the resulting metric according to the defining equation~\eqref{eq: lapse function}. The collapse drives the formation of a black hole horizon whose location lies initially at a radius smaller than its classical counterpart, the latter being approximately located at $r_h/l_{Pl} = 2 m(v)/m_{Pl}$. In fact, the classical Schwarzschild spacetime at the end of the collapse is reproduced well for sufficiently large masses of the final configuration, and only outside of the Planckian region, $r\gg l_{Pl}$. 
In the next section,  Sect.~\ref{sect:Solutionsstatic}, we will analyse possible outcomes of our collapse model and provide an analytical explanation for the origin of the oscillations by studying the static limit of the partial differential equation~\eqref{eq: G infinity differential equation rough}. 

\section{Static spacetimes at the end of the collapse}\label{sect:Solutionsstatic}

A key aspect of classical gravitational collapse models, including the VKP model considered here, is the formation of an event horizon and a spacetime singularity after a finite amount of time. In spherically symmetric settings, the metric at the end of the collapse is the static Schwarzschild geometry and contains a curvature singularity at $r=0$. In terms of the Kretschmann scalar the degree of divergence of the final static configuration is $R_{\mu\nu\rho\sigma} R^{\mu\nu\rho\sigma} \propto {r^{-6}}$. On the other hand, for the quantum-corrected VKP spacetime with the effective gravitational coupling determined by~\eqref{eq: G infinity}, a weakening of the curvature singularity is expected due to the anti-screening character of gravity encoded in Eq.~\eqref{eq: G running}. Within the VKP model, the system is static for advanced times $v>\overline{v}$, as the mass function reaches the constant value $m(v)=m$. In such a static limit the effective Newton coupling becomes independent of the advanced time $v$, $G_\infty = G_\infty(r)$, and the energy density $\mu_\infty$ in Eq.~\eqref{eq: cutoff identification} vanishes. All together, the static limit of the effective Newton coupling is defined by the differential equation~\eqref{eq: G infinity differential equation} with all advanced time derivatives set to zero, and it describes the final spacetime configuration at the end of the gravitational collapse.

To investigate the properties of the resulting static spacetime, in this section we first study the analytical properties of the effective Newton coupling in two opposite limits: in the small radii regime, close to the classical singularity, and in the large distance limit. In the latter the solution displays the same damped oscillations appearing in the collapse phase. Neglecting such oscillations, we will find a function that interpolates between the small- and large-radii behaviors. This interpolating function will provide us with the starting point to study the evaporation phase, which is the focus of the next section.

\subsection{Analytical solution close to the classical singularity}\label{subsubsec: Solutions static fixed-point}

In the following we study the outcome of the quantum-corrected VKP model for small radii. Neglecting for a moment the evaporation effects, at the end of the collapse the effective Newton coupling $G_{\infty}$ will be a function of the radial coordinate only, governed by the differential equation~\eqref{eq: G infinity differential equation} with constant ADM mass $m(v)=m$. Focusing on the small-$r$ region, corresponding to the UV fixed point regime, the RG scale dependence of the dimensionful Newton coupling is given by $G(k) \simeq {g_*}{k^{-2}}$. Since the fixed point regime is reached for $k^2\gg m_{Pl}^2/\omega $ and $\omega = 1/g_* \sim \order{1}$ according to FRG computations, this scaling can be obtained by neglecting the $1$ in the denominator of Eq.~\eqref{eq: G running}. Accordingly, we can study the static spacetime solutions resulting from the collapse, and in the proximity of the classical singularity, by setting $m(v)=m$ (static limit) and by neglecting the $1$ in the denominator of Eq.~\eqref{eq: G infinity differential equation} (fixed point regime). In these limits the effective Newton coupling $G_\infty=G_\infty(r)$ is  completely determined by the ordinary differential equation
\begin{equation}\label{eq: G infinity differential equation static fixed-point}
G_0 \omega m  \qty(4 r G_\infty '' + 8 G_\infty ')G_\infty  - 3 G_0 r^2 = 0\,.
\end{equation}
As we are interested in determining the leading-order scaling of $G_{\infty}(r)$ in the proximity of the would-be singularity, we assume that $G_{\infty}(r)\sim C\,r^n$ close to $r=0$ and determine the parameters $(C,n)$ by inserting this power law ansatz in Eq.~\eqref{eq: G infinity differential equation static fixed-point}. Following this approach we find
\begin{equation}\label{eq: G infinity fixed point static solution}
G_\infty(r)  =  \frac{1}{\sqrt{5 \omega m}}r^{3/2}+\mathcal{O}(r^3)\,.
\end{equation}
Let us stress that Eq.~\eqref{eq: G infinity fixed point static solution} is expected to approximate the effective gravitational coupling at the end of the collapse and at sufficiently small radial distances, $r/r_{Pl} \ll 1$. The $r^{3/2}$-scaling implies that at the origin the effective Newton coupling goes to zero. A positive exponent for the leading power in a series expansion around the origin is consistent with previous results on RG improved black holes in spherical symmetry, e.g.~\cite{Bonanno:2000ep}. In particular, the specific exponent $3/2$ was also found in~\cite{Pawlowski:2018swz}. Using the expression~\eqref{eq: G infinity fixed point static solution} for the Newton coupling in the lapse function of the classical Schwarzschild spacetime,
\begin{equation}\label{eq: f infinity}
f_\infty(r) = 1 - \frac{2 m G_\infty(r)}{r}  \simeq 1 - \frac{2 m}{r}\frac{r^{3/2}}{\sqrt{5 \omega m}}\,,\quad \mathrm{for}\,\, r\ll l_{Pl} \,,
\end{equation}
allows us to investigate properties of the geometry close to the origin. In contrast to the classical solution, the lapse function is regular and takes the value $f_{\infty}=1$ at $r=0$, as a consequence of the vanishing effective Newton coupling in the limit $r\to0$. However, the regularity of the metric at the origin does not imply a curvature singularity-free de Sitter core. Indeed, the Kretschmann scalar of the quantum-corrected VKP model at the end of the collapse becomes
\begin{equation}\label{eq: Kretschmann new}
R_{\mu\nu\rho\sigma}R^{\mu\nu\rho\sigma} \propto \frac{1}{r^3}\,.
\end{equation}
It diverges due to the divergent metric derivative at the origin. Nonetheless, the degree of divergence is lowered compared to the classical singularity and reproduces the exponent found in~\cite{Pawlowski:2018swz} using an alternative cutoff scheme, but a similar self-consistent approach. In summary, the anti-screening character of the gravitational force at high energies reduces the strength of the curvature singularity in comparison to the classical Vaidya model. In previous studies it was shown that such an anti-screening effect might in certain cases even lead to singularity resolution~\cite{Bonanno:1998ye,Bonanno:2000ep,Bonanno:2006eu,Torres:2014gta,Torres:2017ygl}, cf.~also~\cite{Adeifeoba:2018ydh} for an analysis of necessary conditions.
To investigate global properties of the static solutions, an approximate scale dependence of the Newton coupling must be derived from~\eqref{eq: G running}. This is the focus of the next section.

\subsection{Analytical solution at large distances and interpolating function}\label{subsubsec: Solutions static large r}

\begin{figure}[t]
	\centering
	\includegraphics[width=0.47\textwidth]{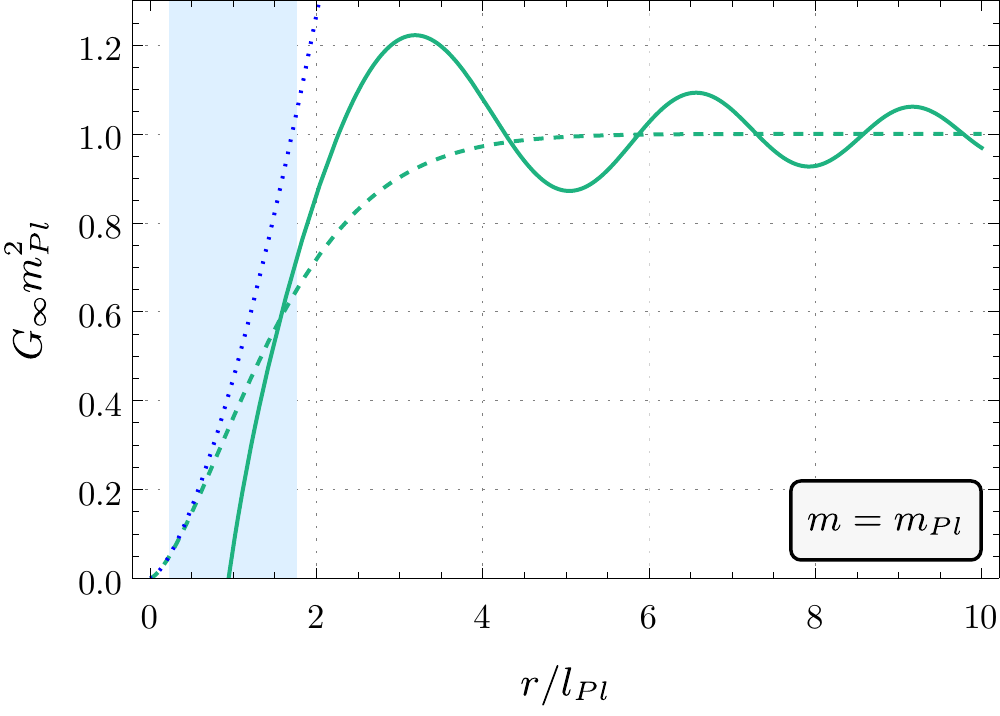}
	\hfill
	\includegraphics[width=0.47\textwidth]{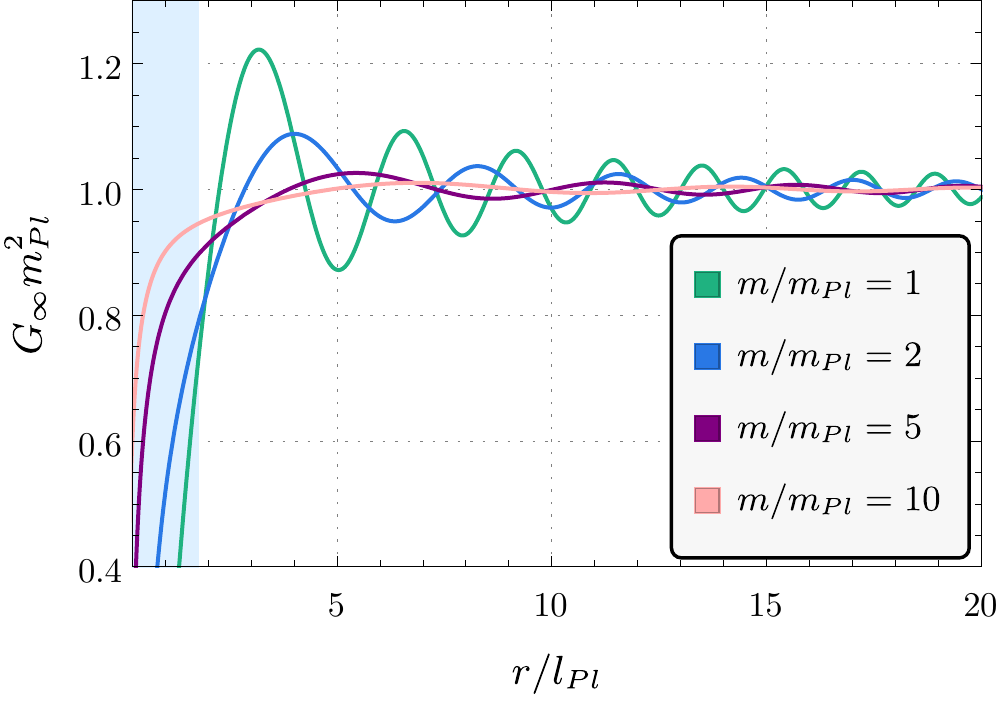}
	\caption{\label{fig: G infinity static}Effective Newton coupling $G_\infty$ as a function of the radial coordinate in the static limit at the end of the collapse, for $\omega\equiv1$. The left panel shows static solutions to Eq.~\eqref{eq: G infinity differential equation} in different approximations and for $m=m_{Pl}$. Below the Planck scale, $r \lesssim l_{Pl}$, the effective Newton coupling is approximated by the solution~\eqref{eq: G infinity fixed point static solution} to the equations in the fixed point regime, and scales as $\sim r^{3/2}$ (dotted line). The solid line displays the analytic solution~\eqref{eq: G infinity large r ansatz} for large radii and is characterized by damped Airy functions of the form~\eqref{eq: Airy F} where we set $c_1,c_2\equiv 1/m$. The blue region is where the transition between these two analytical solutions~\eqref{eq: G infinity fixed point static solution} and~\eqref{eq: G infinity large r ansatz} should occur. Finally, the dashed line shows the exponential function~\eqref{eq: approximate static solution} which smoothly interpolates between the analytic solution in the UV and the Newton's constant $G_0 = m_{pl}^{-2}$ in the IR, and solves Eq.~\eqref{eq: G infinity differential equation} in the static limit and in the special case where the amplitude of the oscillations vanishes.
	The right panel depicts the analytic solution~\eqref{eq: G infinity large r ansatz} with the function $F(r)$ specified by Eq.~\eqref{eq: Airy F} at large radii for different black hole masses. At a given radius, the amplitudes of the oscillations decrease, whereas their wavelengths increase for growing mass parameter $m$. All solutions are valid at large radii, and are not expected to provide a good approximation in the blue region where the transition to the scaling solution~\eqref{eq: G infinity fixed point static solution} occurs.}
\end{figure}

The scope of this section is to determine static analytic solutions at large distances, complementing the analytic solution~\eqref{eq: f infinity} found in the previous section, which instead describes the endpoint of the VKP collapse for small radii. To this end, we need to solve the differential equation~\eqref{eq: G infinity differential equation} with constant mass function $m(v) =m$ and for large radii. Setting $m(v) =m$, the differential equation~\eqref{eq: G infinity differential equation} simplifies to
\begin{equation}\label{eq: G infinity differential equation static}
\qty(G_0 \omega m \qty(4 r G_\infty '' + 8  G_\infty ')   + 3 r^2)G_\infty - 3 G_0  r^2 = 0\,.
\end{equation}
Moreover, at radii $r/l_{Pl}\gg 1$, we can make the ansatz
\begin{equation}\label{eq: G infinity large r ansatz}
G_\infty(r) = G_0\qty(1 - \frac{F(r)}{r})\,,
\end{equation}
where $\abs{F(r)/r}\ll 1$ at large $r$ and  $\abs{F(r)/r}\to 0$ as $r\to \infty$, such that the classical lapse function is recovered at infinity. Inserting the ansatz~\eqref{eq: G infinity large r ansatz} into the differential equation~\eqref{eq: G infinity differential equation static} leads to
\begin{equation}
 4 G_0^{2} m \omega \qty(F(r)-r)F''(r) - 3 r^2 F(r) = 0\,.
\end{equation}
Using that $|F(r)|\ll r$ at large distances, the previous equation reduces to a Stokes differential equation
\begin{equation}
F''(r) + \frac{3}{4 G_0^2 m \omega} r F(r) = 0\,.   
\end{equation}
Solutions are linear combinations of Airy functions in the form 
\begin{equation}\label{eq: Airy F}
F(r) = \mathfrak{Re} \big[c_1 \text{Ai}(a(m,\omega) r) + c_2 \text{Bi}(a(m,\omega)r)\big]\,,   
\end{equation}
where $a(m,\omega) = 2^{-2/3}3^{1/3}(-G_0^2 m \omega)^{-1/3} $. The two integration constants are taken to be $c_i \propto 1/m$ on dimensional grounds (see also~\cite{Carballo-Rubio:2018pmi}). The left panel of Fig.~\ref{fig: G infinity static} shows the analytic solution for the effective Newton coupling at large $r$ according to~\eqref{eq: G infinity large r ansatz} with the function~$F(r)$ given in Eq.~\eqref{eq: Airy F}, together with the analytic solution~\eqref{eq: G infinity fixed point static solution} valid at small radii, for a mass parameter corresponding to one Planck mass, $m=m_{Pl}$. The power-law behavior in the UV remains valid up to approximately one Planck length $r\approx l_{Pl}$ away from the origin. Beyond the transition at the Planck scale where no analytic solution to the differential equation~\eqref{eq: G infinity differential equation static} is available (corresponding to the blue region in Fig.~\ref{fig: G infinity static}), the analytic solution~\eqref{eq: G infinity large r ansatz} characterized by damped Airy functions~\eqref{eq: Airy F} takes over. The presence of Airy functions causes characteristic oscillations around the classical value $G_0 = m_{Pl}^{-2}$ with decaying amplitude and wavelength at increasing radii. In the limit $r\to \infty$ the amplitude of the oscillations goes to zero, such that the classical lapse function is recovered. In particular, since the effective Newton coupling approaches the observed value of Newton's constant in the large-distance limit, the resulting spacetimes are asymptotically flat. In the right panel of Fig.~\ref{fig: G infinity static} the analytic solution~\eqref{eq: G infinity differential equation static} at large radii is displayed for different masses. At a given radius $r$, the amplitudes of the oscillations decrease, whereas their wavelengths increase as the mass parameter $m$ grows. In particular, for astrophysical black holes the mass is $m/m_{Pl} \approx m_\odot /m_{Pl} \approx 10^{38}$ and thus the amplitude of the oscillations becomes tiny and hard to resolve. Accordingly, the energy associated with the inverse wavelength of the oscillations becomes microscopic for large masses. To sum up, the amplitude and wavelength of the oscillations decreases with both the radial coordinate $r$ and with the black hole mass~$m$, making them negligible for astrophysical black holes.
We have additionally confirmed these findings, which are based on our analytic results, through different numerical methods, such as a direct integration of the second-order differential equation, a transformation to a first-order system, and a shooting with boundary conditions imposed at the origin and at large radii.

Let us now comment on the interpretation of these oscillations of the lapse function. On the one hand, similar oscillation patterns were found in certain models of quadratic gravity~\cite{Bonanno:2013dja,Bonanno:2019rsq} (where however the amplitude of the oscillations does not decrease by increasing $r$, and the period does not increase either), in higher-derivative gravity with specific non-local form factors~\cite{Zhang:2014bea}, and in the context of corpuscolar gravity~\cite{Giusti:2021shf,Casadio:2021eio,Casadio:2022ndh}. On the other hand, the RG improvement procedure was originally introduced as a way to explore the leading effects of operators occurring at higher order in the expansion of the effective action. In particular, operators quadratic in the curvature will appear naturally beyond the Einstein-Hilbert truncation. Reproducing solutions to a quadratic action with non-local form factors may therefore be viewed as an indication that the results of the iterative RG improvement coupled with the decoupling mechanism are consistent.

Next, we need to determine an analytic approximation to the full static solution. If the oscillations on top of the effective Newton coupling are neglected, we find that the analytic power-law solution~\eqref{eq: G infinity fixed point static solution} at the origin and the classical constant Newton coupling at large~$r$ are smoothly connected by the approximate interpolating function
\begin{equation}\label{eq: approximate static solution}
G_\infty^{int}(r) = G_0\qty(1 - e^{-\frac{r^{3/2}}{\sqrt{5 \omega r_h/2}\,l_{Pl}}})\,, 
\end{equation}
with $r_h = 2 m G_0$. 
This can be seen by computing the next-to-leading order corrections to Eq.~\eqref{eq: G infinity fixed point static solution} and comparing them with the expansion of various possible interpolating functions, perhaps inspired by the most commonly studied black holes beyond GR. Specifically, the corrections to Eq.~\eqref{eq: G infinity fixed point static solution} read
\begin{equation}\label{eq:expanded-G-nextorders}
	G_\infty(r)=\frac{r^{3/2}}{\sqrt{5m\omega}}-\frac{r^3}{21 G_0 m \omega }+\frac{25  r^{9/2}}{16758 \sqrt{5} G_0^2 (m \omega) ^{3/2}}+\mathcal{O}(r^{11/2})
\end{equation}
and we verified that the above next-to-leading order coefficients do not match those of Bardeen-like or Hayward-inspired solutions. One of the reasons is that they would introduce, e.g., $r^4$- or $r^{5/2}$-corrections that are absent in Eq.~\eqref{eq:expanded-G-nextorders}. In contrast, the expansion of Eq.~\eqref{eq: approximate static solution} (inspired by the Dymnikova solution) is
\begin{equation} 
	G_\infty^{int}(r)=\frac{r^{3/2}}{\sqrt{5m\omega}}-\frac{r^3}{10 G_0 m \omega }+\frac{r^{9/2}}{30 \sqrt{5} G_0^2 (m \omega) ^{3/2}}+\mathcal{O}(r^{11/2})
\end{equation}
and offers a better approximation to the expansion~\eqref{eq:expanded-G-nextorders}.

The interpolating function~\eqref{eq: approximate static solution} is shown in the left panel of~Fig.~\ref{fig: G infinity static}.
In the limit of large masses or large radii the exponential becomes negligible, and deviations from the classical Schwarzschild solution are strongly suppressed. The exponential nature of the interpolating function seems to be a feature of the self-consistent approach, which in the static case leads to a Dymnikova solution, cf.~\cite{Platania:2019kyx}, corresponding to an effective Newton coupling of the type~\eqref{eq: approximate static solution}, with characteristic scaling $\sim r^3$ close to the origin, in place of $\sim r^{3/2}$. The Dymnikova scaling is physically more appealing, as it makes curvature invariants finite at $r=0$. This is to be contrasted with our case, where the characteristic scaling $\sim r^{3/2}$ of the effective Newton coupling is not strong enough to remove the singularity, although it makes it weaker, cf.~\eqref{eq: Kretschmann new}. This result is not surprising: even at the level of a one-step RG improvement, adding quantum corrections to the static Schwarzschild solution leads to singularity resolution~\cite{Bonanno:2000ep}, while, starting from a dynamical spacetime, the dynamics of the quantum-corrected gravitational collapse typically lead to black holes with gravitationally weak (or integrable~\cite{Lukash:2011hd}) singularities~\cite{Bonanno:2017zen}. Replacing the one-step RG improvement with the self-consistent procedure in~\cite{Platania:2019kyx} does not change this intriguing result. Yet, in contrast to the one-step RG improvement, self-consistency favors the appearance of exponential lapse functions. Such an exponential behavior is a highly desirable feature, as it gives hopes that the corresponding spacetime can come from a principle of least action in quantum gravity~\cite{Knorr:2022kqp}. In particular, it is conceivable that these spacetimes characterized by exponential lapse functions could stem from an effective action of the type~\eqref{eq:eff-action} with exponential form factors. Notably, this resonates with the findings in~\cite{Zhang:2014bea}, where it was shown that quadratic effective actions with exponential form factors lead to damped oscillations resembling those that we have observed. In contrast, the typical polynomial lapse functions obtained from the one-step RG improvement, such as the Bonanno-Reuter metric~\cite{Bonanno:2000ep} and the Hayward black hole~\cite{Hayward:2005gi}, seem to be incompatible with a principle of least action~\cite{Knorr:2022kqp}, making their relation with quantum gravity questionable. 

\section{Dynamics of the evaporation process}\label{sect:evaporation}

In the previous section we obtained an approximate analytical result for the effective Newton coupling at the end of the collapse. According to Eq.~\eqref{eq: approximate static solution}, the resulting static metric is characterized by the approximate lapse function
\begin{equation}\label{eq: lapse function static solution}
f_\infty(r) = 1-\frac{2 m G_0}{r}\qty(1 - e^{-\frac{r^{3/2}}{\sqrt{5 \omega r_h/2}\,l_{Pl}}})\,,
\end{equation}
where we remind the reader that $m$ is the ADM mass measured by an observer at infinity. Although the lapse function~\eqref{eq: lapse function static solution} neglects the oscillations encountered in the previous section, it provides an analytical approximation to the endpoint of the gravitational collapse and sets our starting point to study its evaporation.

In the following we will work under the assumption that the black hole radiation remains thermal until the end of the evaporation. While this is likely not realized due to non-perturbative effects, it is not straightforward to describe the complete evaporation in a non-perturbative fashion. Moreover, this assumption will allow us to compare our case to the classical one of Schwarzschild black holes. 

\subsection{Causal structure and critical mass}\label{subsubsec:alltogether}

The interpolating function~\eqref{eq: approximate static solution} allows us to study the causal structure of the quantum-corrected static spacetime at the end of the collapse, and to determine the approximate location of its horizon(s). While the classical lapse function has a single horizon at $r_h=2G_0 m$, the causal structure of the quantum-corrected spacetime is more complicated and, similarly to other proposed alternatives to Schwarzschild black holes, it depends on the ratio $m/m_{Pl}$. At a critical value $m=m_c$ there is exactly one horizon. For masses below the critical mass there is no horizon and the curvature singularity is naked and timelike. Above the critical mass instead, as is typical for regular black holes, there are two horizons. We note at this point that our construction does not eliminate the problem of mass inflation characterizing most black holes with two horizons, as the lapse function~\eqref{eq: lapse function static solution} is such that the surface gravity at the inner horizon $\kappa_-$ is non-zero~\cite{Carballo-Rubio:2022kad}.

\begin{figure}[t]
\centering
\includegraphics[width=0.6\textwidth]{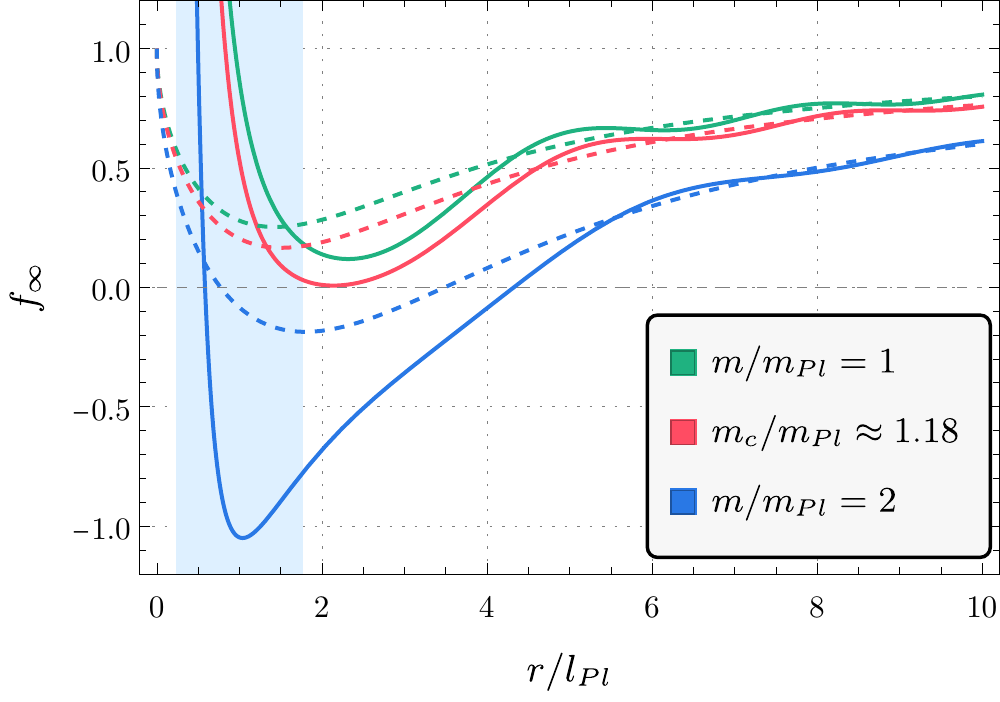}
\caption{\label{fig: lapse function} Lapse function $f_\infty$ of the final static configuration as a function of $r$ for different masses, each depicted with a different color. The analytic solution based on the oscillating effective Newton coupling~\eqref{eq: G infinity large r ansatz} with the function $F(r)$ defined by~\eqref{eq: Airy F} (dashed lines) is valid at large radii, $r/l_{Pl}\gg 1$. In the deep UV instead, for $r\ll l_{Pl}$, the lapse function is approximated by the analytic solution to equation~Eq.~\eqref{eq: G infinity fixed point static solution} (solid lines), and takes the value $f_\infty(0) = 1$ at the origin. The region highlighted in blue is where the transition between these two analytic solutions, which are valid in opposite asymptotic regimes, should occur.}
\end{figure}

The critical ratio $m_c/m_{Pl}$ is expected to be of order one, since no scale other than the Planck mass is included in our physical description. The critical mass parameter can be estimated analytically as follows.
First, we start by determining the condition to have a horizon close to the classical singularity, where the Newton coupling is described by the function~\eqref{eq: G infinity fixed point static solution}. This is done by inserting~\eqref{eq: G infinity fixed point static solution} into~\eqref{eq: f infinity} and searching for zeros of the resulting lapse function. There is one zero at 
\begin{equation}\label{eq: horizon location approximated}
r_h = \frac{5 \omega}{4 m/m_{Pl}} l_{Pl}\,.
\end{equation}
Next, we recall that the fixed point scaling of the Newton coupling is valid only at high energies, i.e.,~at small distances, $r/l_{Pl} \ll 1$. The previous condition is saturated in Eq.~\eqref{eq: horizon location approximated} at $r=r_h$, if the mass parameter is chosen to be $m_c/m_{Pl} \approx \sqrt{5/4} \simeq 1.12$. This derivation however is valid only if the horizon lies in the region where the analytic approximation for $G_\infty$ based on the fixed point solution is adequate and we must verify this assumption a posteriori. It turns out that $m_c/m_{Pl} \approx \sqrt{5/4}$ can only provide a rough estimate for the critical mass $m_c$. In fact, the horizon location for this value of the mass parameter would be at $r\approx 2 G_0 m$, cf.~Fig.~\ref{fig: lapse function}, which is outside the regime of validity of Eq.~\eqref{eq: G infinity fixed point static solution} but within the same order of magnitude. A numerical analysis utilizing the analytical approximation~\eqref{eq: G infinity large r ansatz} for the lapse function at large radii shows that the correct value for the critical mass lies slightly above the  analytical one derived above, and is $m_c/m_{Pl} \approx 1.18$, cf.~Fig.~\ref{fig: lapse function}.

\begin{figure}[t]
\centering
\includegraphics[width=0.6\textwidth]{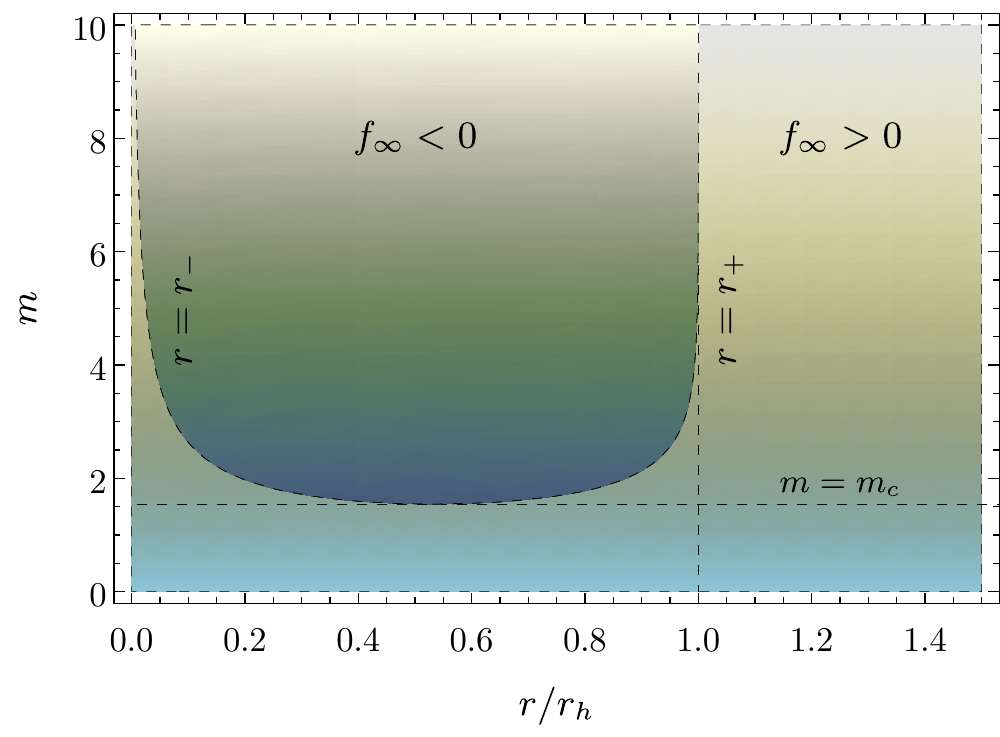}
\caption{\label{fig: f infinity} Density plot of the lapse function $f_\infty(r)$, highlighting its positivity for increasing values of the mass parameter~$m$ and as a function of the radial coordinates in units of $r_h = 2 G_0 m$. In the figure $r_-$ and $r_+$ denote the inner and outer horizon, respectively. For $r> r_h$ or $m<m_c$ the lapse function $f_\infty(r)$ is strictly positive. The causal structure is instead non-trivial for $m\geq m_c$, where $f_\infty(r)$ can also be negative or vanish. Specifically, the lapse function is negative between the two horizons, positive outside, and vanishes on the boundary. Thus, for masses $m/m_{Pl}$ greater than, equal or less than the critical value $m_c/m_{Pl}$, the spacetime has two, one or no horizon(s) respectively.}
\end{figure}

Finally, when neglecting the oscillations, i.e., when considering the interpolating lapse function~\eqref{eq: lapse function static solution} as a starting point, the qualitative causal structure is similar: depending on the value of $m$, the spacetime exhibits two, one or no horizons, cf. Fig.~\ref{fig: f infinity}. The location $r_+$ of the outer black hole horizon approximates the location of the classical Schwarzschild radius at $r_h = 2 G_0 m$ as the mass is increased. At the same time, increasing $m$ the inner Cauchy horizon $r_-$ moves closer to the origin in units of $G_0 m$. At the critical value $m_c$ both horizons coincide, while spacetimes characterized by smaller masses have no horizon. Finally, if the black hole mass $m$ is below the critical value $m_c$, the spacetime is horizon-free. In this case however the critical mass is $m_c \approx 1.55 \, m_{Pl}$. The difference with the one previously discussed stems from neglecting the oscillations of the lapse function, as is clear from Fig.~\ref{fig: lapse function}. As the specific position of the horizon does not impact the qualitative aspects of the evaporation process, and since we do not have a full solution featuring both the oscillations~\eqref{eq: G infinity large r ansatz} at large radii and the correct $\sim r^{3/2}$ scaling at short distances, we will neglect the oscillations and we will use the interpolating lapse function~\eqref{eq: lapse function static solution} as a starting point to study the evaporation process of the corresponding black hole. 

\subsection{Evaporation process}

For $m > m_c$ the lapse function $f_\infty$ exhibits a simple zero at $r=r_+$. In particular, its derivative is non-negative for $r\geq r_+$ and its value increases monotonically from zero at $r=r_+$ to one at infinity. We may thus associate a temperature to this black hole configuration by following Hawking's analysis of black hole radiance~\cite{Hawking:1975vcx} in the language of Euclidean path integrals and thermal Green's functions~\cite{Gibbons:1976pt,Gibbons:1976ue,Hawking:1978jz}. 
To this end, let us consider a static spherically symmetric spacetime of the form
\begin{equation}\label{eq: Vaidya metric}
\dd{s^2} = -f(r)\dd{t^2} + f(r)^{-1} \dd{r^2} + r^2\dd{\Omega^2}\,.
\end{equation}
A positive definite Euclidean metric can be defined by performing a Wick rotation, i.e., by complexifying the time coordinate, $t\to i\tau$. Expanding the lapse function in a Taylor series in the near-horizon region it can be shown that to first order the metric locally describes a Rindler space. A coordinate transformation $(\tau, r) \to (\phi, \rho)$, where $\phi = \abs{f'(r_+)}\tau/2$ and $\rho^2 = 4 (r-r_+)/f'(r_+)$, allows us to write the metric in the neighborhood of the horizon as
\begin{equation}
\dd{{s_E}^2} = \dd{\rho^2} + \rho^2 \dd{\phi^2} + r_+^2\dd{\Omega^2}\,.
\end{equation}
By requiring smoothness of the metric, one is led to identify $\phi$ with an angle variable having period $2\pi$ and to restrict the range of possible values of the radial variable to $r> r_+$. In this case the first two terms in the Euclidean metric correspond to the line element of a 2-dimensional flat plane written in polar coordinates $(\phi, \rho)$. The resulting manifold is a Euclidean black hole with topology $\mathbb{R}^2 \times S^2$. 

\begin{figure}[t]
\centering
\includegraphics[width=0.47\textwidth]{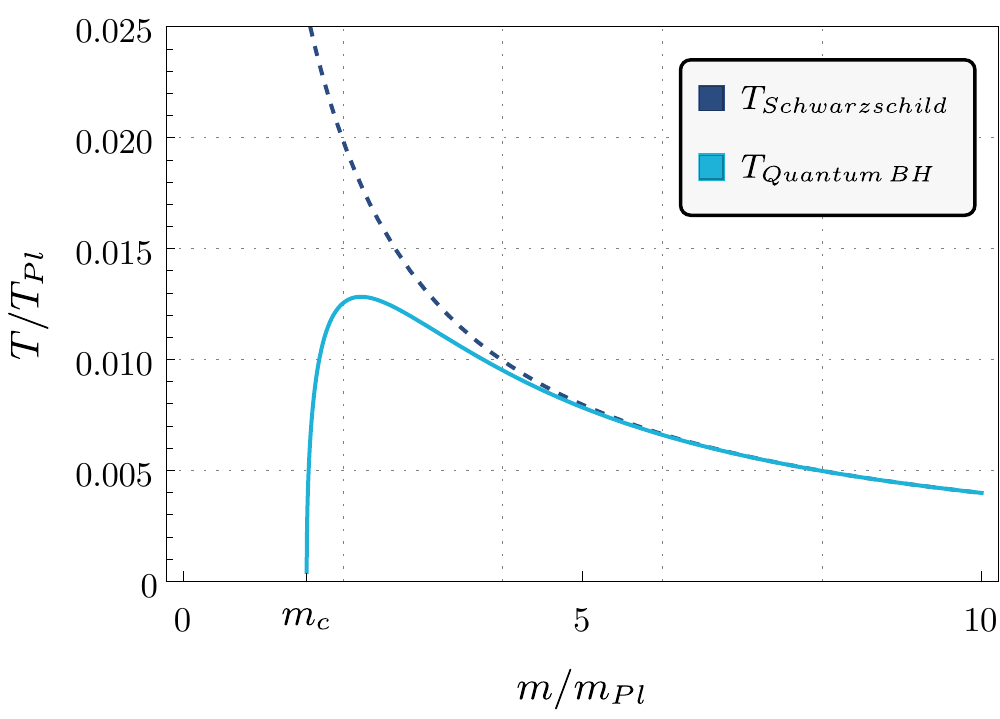}
\hfill
\includegraphics[width=0.47\textwidth]{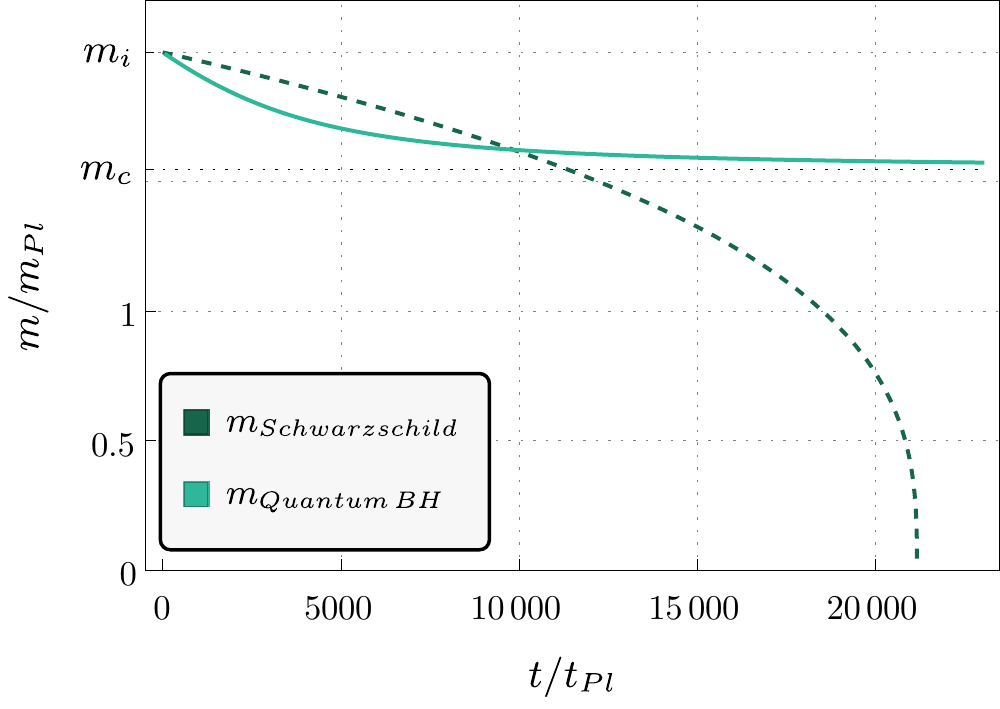}
\caption{\label{fig: temperature and evaporation} Temperature and mass of an evaporating classical and quantum black hole. All quantities are in Planck units. The left panel depicts the temperature of the quantum black hole (solid blue line) compared to the classical Hawking temperature of a Schwarzschild black hole (dashed dark blue line). The temperature is displayed as a function of the black hole mass. In the classical evaporation process, the black hole becomes hotter and hotter, leading to a complete evaporation which eventually leaves the classical singularity naked after a finite amount of time. In the quantum version, the temperature at first increases as the mass decreases. However, in contrast to the classical case, it reaches a maximum and then slowly goes to zero. The right panel shows the black hole mass as a function of the proper time measured by an observer at infinity. The function $m(t)$ is determined by solving Eq.~\eqref{eq:temperaturevariation} numerically, with the initial value $m_i \equiv m(0)$ set to $m_i/m_{Pl}=2 >m_c$ and with $\sigma\equiv 1$.  The dashed dark green line corresponds to the classical case, and shows that the evaporation process occurs in a finite amount of time. By contrast, in the quantum-corrected model (solid green line), a black hole with initial mass $m_i$ requires an infinite amount of proper time to convert the mass $(m_i-m_c)$ into Hawking radiation, eventually leading to a black hole remnant with mass $m_c$ (dotted black line).}
\end{figure}

The periodicity of $\phi$ translates into one of $\tau$, i.e.~$\tau \to \tau + \beta$ with period $\beta = 4\pi / f'(r_+) $. If quantized matter fields are considered on the Euclidean black hole background, their Green's functions become thermal with the temperature determined by the inverse of the parameter $\beta$,
\begin{equation}\label{eq: Temperature}
T_{BH} = \frac{1}{\beta} = \frac{f'(r_+)}{4\pi}\,.   
\end{equation}
For a Schwarzschild black hole this temperature reproduces the well-known result due to Hawking~\cite{Hawking:1975vcx},
\begin{equation}\label{eq: Hawking temperature}
T_{Schwarzschild} = \frac{1}{8\pi G_0 m}\,.
\end{equation}
In order to apply the previous formula to our case, starting from the configuration with two horizons, one has to identify the location of the outer horizon $r_+$. The latter is given by largest positive zero of the lapse function~\eqref{eq: lapse function static solution}, see~Fig.~\ref{fig: f infinity}. We determine this root numerically for varying mass and insert the result into~\eqref{eq: Temperature}. Thereby we arrive at the temperature as a function of $m$, shown in the left panel of Fig.~\ref{fig: temperature and evaporation}. For large masses the spacetime is well approximated by the Schwarzschild solution. As a consequence, in this limit the temperature of the quantum black hole reduces to the Hawking temperature~\eqref{eq: Hawking temperature}. Lowering the mass, deviations between the classical and quantum spacetime become significant, with the quantum-corrected temperature always lying below the semi-classical one. While the latter diverges as $\propto 1/m$ for small $m$, when lowering the mass of the quantum black hole its temperature reaches a maximum and subsequently falls down to zero. This happens when the two horizons coincide, i.e., when the black hole mass $m$ has reached the critical value~$m_c$. Initial configurations characterized by a smaller mass parameter have no horizon and thus the derivation of~\eqref{eq: Temperature} does not apply. 

Our results are in remarkable agreement with the RG improved spacetimes studied in~\cite{Bonanno:2000ep}, whereby a cutoff function is constructed from the radial proper distance of an observer to the center: after reaching a maximum temperature, the quantum black hole begins to cool down. The evaporation process comes to an end when its mass is lowered to $m=m_c=\order{m_{Pl}}$. The critical mass therefore represents a final state of evaporation, leaving behind a Planck-size black hole remnant.

We now evaluate how much time is needed for the evaporation of a black hole from an initial mass $m_i$ to its final value $m_f$. The mass loss per unit proper time measured by an observer is given by Stefan-Boltzmann's law
\begin{equation}\label{eq:temperaturevariation}
\dot{m} = - \sigma A(m) T_{BH}^4(m)\,,
\end{equation}
where a dot denotes differentiation with respect to the proper time $t$, $\sigma$ is a constant and $A(m)=4\pi r_+ ^2$ is the area of the outer horizon. For a Schwarzschild black hole the radiation power decreases as $\propto m^{-2}$, which leads to a finite amount of time $\propto {m_i}^3$ for the complete evaporation from $m_i\to 0$ to happen, as shown in the right panel of Fig.~\ref{fig: temperature and evaporation}.
The situation is notably different in the quantum case. Starting from an initial value $m_i>m_c$, the critical final value $m_c$ is reached only asymptotically, at infinitely late times. This can be explained as follows. As the quantum black hole evaporates it eventually reduces its mass to the value associated with the maximum temperature peak, displayed in Fig.~\ref{fig: temperature and evaporation}. Thereafter the cooling process begins and the temperature gradient becomes negative. When the temperature is close to zero, the mass change per time---which obeys a $T^4$-behaviour according to Stefan-Boltzmann's law---becomes tiny. At this stage the black hole cannot radiate away power efficiently anymore. In particular, it is impossible to reach the final stage of evaporation. This result is consistently interpreted in view of the third law of black hole thermodynamics, according to which a zero surface gravity cannot be achieved in a physical process, as has already been observed in~\cite{Bonanno:2000ep}.

\begin{figure}[t!]
\centering 
\includegraphics[width=0.47\textwidth]{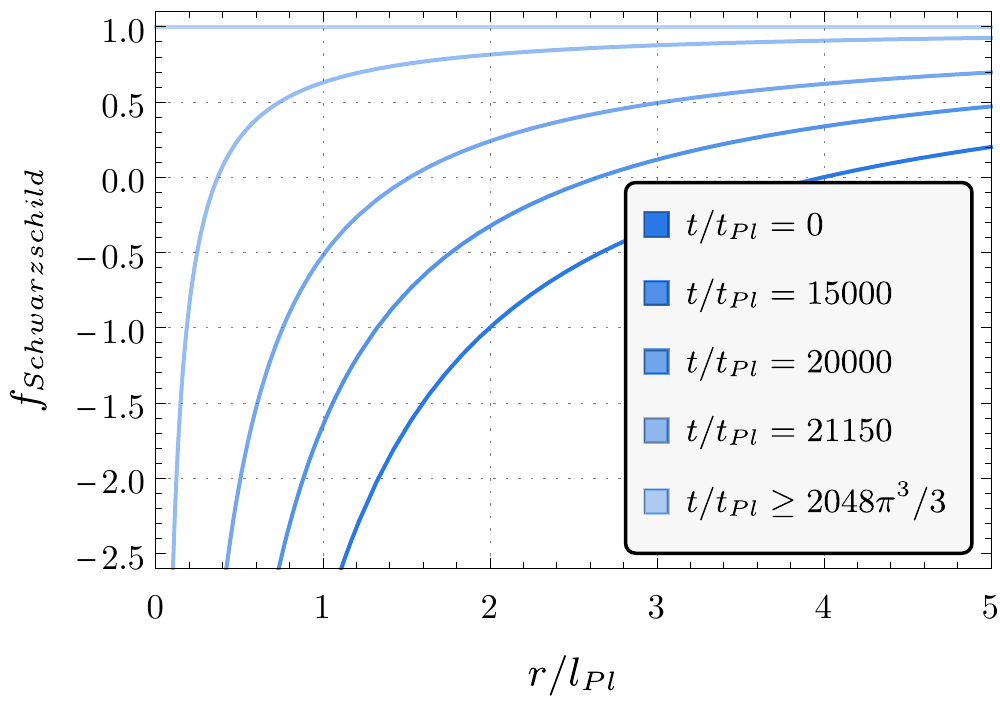}
\hfill
\includegraphics[width=0.47\textwidth]{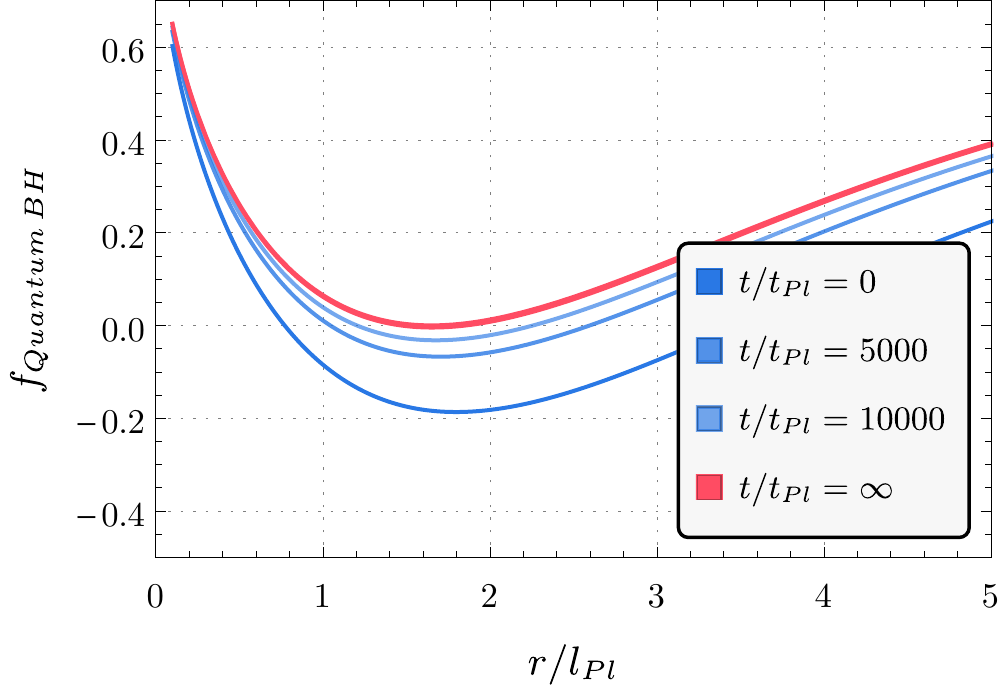}
\caption{\label{fig: evaporation lapse function}Schwarzschild (left panel) and quantum-corrected lapse function (right panel) at different times. Classically, the Hawking temperature increases monotonically as the black hole mass is converted into Hawking quanta. The evaporation thus continues for a finite amount of time $\Delta t=2048 \pi^3/3 t_{Pl}$ until the mass $m$ vanishes and the classical black hole reduces to a Minkowski spacetime with a naked singularity at $r=0$. The corresponding lapse function is one everywhere except at the origin, where it diverges. In the quantum model, evaporation takes an infinite amount of time and a black hole remnant with mass $m_c\sim m_{Pl}$ (solid red line) is formed asymptotically.}
\end{figure}

The time dependence of the metric is obtained by plugging the time-dependent mass function $m(t)$ into the Vaidya lapse function~\eqref{eq: lapse function}. Fig.~\ref{fig: evaporation lapse function} shows the evaporation of a Schwarzschild black hole compared to the time evolution of its quantum counterpart. A Schwarzschild black hole evaporates completely within a time $t\propto m_i ^3$, leaving behind  empty Minkowski space, modulo a naked singularity. By contrast, a quantum black hole gradually approaches the critical configuration for which the inner and outer horizon coincide. Following our previous considerations, however, it will take infinitely long to get there.

We finally comment on the applicability of the entropic arguments against remnants~\cite{Bekenstein:1993dz,Susskind:1995da} to our case. In principle, if one assumes that only asymptotic states matter and that black holes are uncharged, there are at least two ways to avoid the conclusions of~\cite{Bekenstein:1993dz,Susskind:1995da}: \textit{(i)} remnants are unstable and only asymptotic states matter; \textit{(ii)} remnants are stable but there is no degeneracy. Since in our case all remnants have the same mass and in principle the same entropy (provided that information can escape, e.g., via non-thermal processes at the end of the evaporation), this might be enough to remove the dangerous degeneracy typically associated with remnants. This solution might however not work in the case of charged black holes (which should be carefully checked separately), as the presence of a charge would likely not allow having remnants of the same mass. This would likely re-introduce the degeneracy and is one of the strongest arguments supporting the idea that there can be no global symmetries in quantum gravity~\cite{Agmon:2022thq}. While the ``no global symmetry'' conjecture has not been tested within asymptotic safety\footnote{See~\cite{Basile:2021krr} for a preliminary comparison of the string and asymptotic safety landscapes.}, there is currently no evidence supporting a conflict between global symmetries and the asymptotic safety condition. If more refined calculations would confirm this conclusion, alternative explanations for degeneracy avoidance ought to be sought after. A possibility would be that asymptotic safety is a low-energy approximation to a more fundamental theory forbidding global symmetries (e.g., string theory~\cite{deAlwis:2019aud,Basile:2021euh}). Alternatively, the degeneracy might be avoided if black holes have quantum hair~\cite{Calmet:2021stu}, or if the evaporation process becomes highly non-perturbative in its last stages, in a way that allows all information to escape. These are exciting possibilities and intriguing research avenues, to which we hope to contribute in the future.

\section{Conclusions}\label{sect:conclu}

A key challenge of quantum gravity is to derive spacetimes whose properties and dynamics are valid at all resolution scales. Such dynamical solutions are expected to emerge from a principle of least action, in which the classical action is replaced by its quantum (or ``effective'') counterpart. Yet, determining such an effective action as well as finding solutions to the corresponding quantum field equations is technically extremely involved. One should first evaluate the gravitational path integral or, equivalently, solve the RG equations of a scale-dependent version of the effective action~\cite{Dupuis:2020fhh}. By taking its infrared limit, all quantum fluctuations are integrated out and the scale-dependent effective action reduces to the standard quantum effective action.

As a way to circumvent these technical challenges, in the past decades studies of quantum gravity phenomenology in the context of asymptotically safe gravity have strongly relied on the use of ``RG improvement''~\cite{Coleman1973:rcssb,Migdal:1973si,Adler:1982jr,Dittrich:1985yb}. The latter was originally devised in the context of quantum field theory to provide insights on the quantum dynamics while avoiding the complex procedures of solving RG equations or computing quantum loops in perturbation theory. Its necessary ingredients are an action, the beta functions governing the scale dependence of its couplings, and a functional relation between the RG scale and the characteristic energy of a given phenomenon, e.g., the center of mass energy in a scattering process. Although the use of RG improvement in quantum field theory has been incredibly successful, its application to gravity is subject to several ambiguities (see, e.g.,~\cite{Platania:2020lqb} for a summary), making its connection to the asymptotic safety program unclear. In particular, the lack of a clear recipe to relate the RG scale with the variety of competing physical energy scales involved in gravitational phenomena is one of its most severe problems.

In this work we put forth a method to address this issue and to determine some of the leading-order quantum corrections to classical spacetimes. Our strategy relies on the so-called decoupling mechanism~\cite{Reuter:2003ca}: when a system is characterized by one or more physical infrared scales, their combination can overcome the regulator term implementing the shell-by-shell integration of fast-fluctuating modes in the path integral, thus slowing down the flow of the scale-dependent effective action. At the ``decoupling scale''---the critical scale below which the flow freezes out---the scale-dependent effective action approximates the quantum effective action. The decoupling mechanism thus provides a short-cut to the effective action and generally grants access to higher-order terms which were not part of the original truncation. In this work we derived a condition to identify the decoupling scale, given an ansatz for the action, and subsequently exploited this condition to study the dynamics of quantum-corrected black hole spacetimes in asymptotic safety, starting from the Einstein-Hilbert truncation.

Our results are remarkably promising. On the one hand, they are in qualitative agreement with previous studies based on the RG improvement. Specifically:
\textit{(i)} Accounting for the dynamics of a gravitational collapse makes full singularity resolution less straightforward than in static settings. Nevertheless, quantum effects make the singularity gravitationally weaker, in agreement with preliminary indications from first-principle computations~\cite{Bosma:2019aiu}; \textit{(ii)} Black holes can have up to two horizons depending on whether their mass is below, equal, or above a critical Planckian mass scale. Astrophysical black holes would thus be characterized by two horizons and their evaporation would resemble closely the one of known black holes in the literature~\cite{Dymnikova:1992ux,Hayward:2005gi,Bonanno:2006eu}. On the other hand, in our construction we find additional striking features reminiscent of higher-derivative operators with specific non-local form factors. In particular, the lapse function characterizing quantum-corrected black holes decreases exponentially, and displays damped oscillations along the radial direction. Although we started from the Einstein-Hilbert truncation, free oscillations are typical of black holes in local quadratic gravity assuming a specific sign of the Weyl-squared term~\cite{Bonanno:2013dja,Bonanno:2019rsq}. This result is consistent with the expectation that the decoupling mechanism ought to grant access to higher-derivative terms that were not included in the original truncation, and provides encouraging evidence that our construction could lead to results in qualitative agreement with first-principle calculations in quantum gravity. In addition, the damping of the oscillations indicates the presence of non-local form factors in the quadratic part of the effective action. Specifically, given the exponential nature of the dynamical lapse function we derived, one could speculate that these black holes could stem from an effective action with exponential form factors. In turn, this hypothesis is supported by the findings in~\cite{Zhang:2014bea}, where it was shown that exponential form factors in the action yield black holes whose lapse functions oscillate along the radial direction, with a characteristic damped amplitude.

Altogether, the decoupling mechanism provides an intriguing novel avenue to systematically compute leading-order corrections to classical spacetime solutions. While in this work we focused on black holes and we started from a simple ansatz for the action, our construction also applies to cosmological frameworks and can be extended to include higher-derivative terms. Specifically, it is both interesting and necessary to investigate the stability of our findings against the truncation order. This requires going beyond the Einstein-Hilbert truncation, introducing for instance quartic and cubic curvature invariants, and determining the corresponding black-hole solutions from the decoupling mechanism. We hope to tackle these points in future works.

\acknowledgments

The authors would like to thank Niayesh Afshordi, Ivano Basile, Benjamin Knorr, and Nobuyoshi Ohta for interesting discussions, and Benjamin Knorr for very helpful comments on the manuscript. JNB is supported by NSERC grants
awarded to Niayesh Afshordi and Bianca Dittrich. The authors acknowledge support by Perimeter Institute for Theoretical Physics. Research at Perimeter Institute is supported in part by the Government of Canada through the Department of Innovation, Science and Economic Development and by the Province of Ontario through the Ministry of Colleges and Universities. AP also acknowledges Nordita for support within the ``Nordita Distinguished Visitors'' program and for hospitality during the last stages of development of this work. Nordita is supported in part by NordForsk. 

\printbibliography{}

\end{document}